\documentclass[pre,aps,
article,
aps,
superscriptaddress,floats,showpacs,pdftex]{revtex4}
\usepackage{graphicx}
\usepackage{amsmath}
\usepackage{amssymb}
\usepackage[T1]{fontenc}
\usepackage{color}

\begin{document}

\title{Relativistic Mirrors in Laser Plasmas (Analytical Methods)}
\author{S. V. Bulanov$^{1,2}$, T. Zh. Esirkepov$^1$, M. Kando$^1$, J. Koga$^1$
\\
\bigskip
$^1${\small { Kansai Photon Science Institute, JAEA, 8-1-7 Umemidai,
Kizugawa, Kyoto, 619-0215  Japan}}\\
$^2${\small {A. M. Prokhorov Institute of General Physics, Russian Academy of Sciences,\\
Vavilov street, 38,  Moscow, 119991 Russia}}
\\}

\begin{abstract}
Relativistic flying mirrors in plasmas are realized as  thin dense
electron (or electron-ion) layers accelerated by high-intensity
electromagnetic waves to velocities close to the speed of light
in vacuum. The reflection of
an electromagnetic wave from the relativistic mirror results in
its energy and frequency changing. In a counter-propagation
configuration, the frequency of the reflected wave is multiplied
by the factor proportional to the Lorentz factor squared. This
scientific area promises the development of sources of ultrashort
X-ray pulses in the attosecond range. The expected intensity
will reach the level at which the effects predicted by nonlinear
quantum electrodynamics start to play a key role. 
\end{abstract}

\date{\today}
\maketitle

\tableofcontents

\newpage

\section{Introduction} 

\label{intro}
{Sources of electromagnetic radiation with wavelengths in the range from microwave to 
X-ray and gamma-ray are required for a wide variety of problems in fundamental and applied physics.
 
One of the approaches to realize compact sources of hard electromagnetic radiation 
with controllable parameters is based on the relativistic mirror concept. 
Within the framework of this concept an electromagnetic wave reflected from 
a counter-propagating relativistic mirror is compressed in the longitudinal direction 
and its carrier frequency is upshifted \cite{Einstein} 
(see also the review \cite{MIRR} and the literature cited therein).

The question on realization of relativistic mirrors under laboratory 
conditions has been raised repeatedly over a number of years. 
In this regard emerges a question on whether or not it is possible to prepare 
the relativistic mirror of high enough quality for efficient reflection of the light, 
which can move with substantially large velocity for providing significantly 
high frequency increase up to the level corresponding to the photon energy in the X-ray range. 
The answer can be found using the knowledge in the physics of nonlinear processes 
in relativistic laser plasmas. 

One of the ways towards achieving the higher frequency and intensity
is based on the simultaneous laser frequency upshifting and pulse compression. 
It was demonstrated within the relativistic flying mirror (RFM) concept, 
which uses the laser pulse compression, frequency up-shift, and focusing 
by counter-propagating thin electron shells \cite{MIRR, BET-03}. 
The results of corresponding proof-of-principle experiments are presented in Refs. 
\cite{K1, P1, K2} and in Section VI below.

The RFM concept requires substantially high intensity laser radiation interaction with matter. 
Recent progress in laser technology has lead to a dramatic increase of 
laser power and intensity, opening new fields of regimes of laser-matter interaction
\cite{MTB, MaShu-2006, EPS-Stra, ADiP-2012, FUNDASTRO}.
At intensities greater than 10$^{18}$W/cm$^{2}$, the electron quiver energy 
exceeds its rest mass energy.
Under these conditions the laser matter interaction 
can be described by the electrodynamics of continuous media.
In the near future the radiation of ultrashort pulse high power lasers may reach 
the intensity of 10$^{24}$W/cm$^{2}$ and higher \cite{ELIILE}.
In this limit protons also become relativistic, while the electron motion
becomes strongly dissipative due to its intense hard radiation emission,
so that the interaction
should be described within the framework of quantum electrodynamics (QED).

In the relativistic regime of the interaction of a high intensity electromagnetic wave with 
plasmas the nonlinear character of this interaction 
results in the formation of electron 
density modulations, including dense electron bunches and/or 
thin electron layers moving with
relativistic velocity.
An interaction of the same or another electromagnetic wave with 
these relativistic objects can be reduced to the wave reflection from relativistic mirrors,
which is accompanied by the relativistic frequency shift and change of the reflected wave amplitude. 
This is the mechanism of 
the wave intensification and frequency upshifting,
including high order harmonics generation, 
in relativistic collisionless plasma.
This underlies the relativistic oscillating mirror concept \cite{ROM}
and the generation of electromagnetic attosecond pulses \cite{PLAJA98, HOHSD}.
The robustness of this mechanism allowed the 
experimental demonstration of attosecond phase locking of harmonics emitted 
from laser produced plasmas \cite{NOMURA}.
The light frequency up-shifting  and intensification 
have been discussed in the problem of a laser interaction with solid density targets,
which revealed attosecond pulse and high order harmonics generation
\cite{HOHSD, HOHSD1, HOHSD2, PLAJA98}.

We notice a principal difference between two physical situations.
One is the electromagnetic wave scattering by a single relativistic electron, 
which corresponds to Thomson or, in the quantum limit, 
to Compton scattering.
Another is the wave reflection by a dense enough electron 
layer moving with relativistic velocity.
In both cases the maximum frequency upshift is the same being proportional 
to the square of the electron's Lorentz factor. 
However, the number of the emitted photons differs. 
While in the first case the scattered light intensity is 
linearly proportional to the number of electrons,
in the second case the reflected light intensity is 
quadratically proportional to that number. 
In addition, in the first case the scattered radiation is incoherent, 
and in the second case the reflected light is coherent.

These properties are of paramount importance for developing electromagnetic radiation sources 
capable of providing the wave intensity approaching the quantum electrodynamics limits \cite{BET-03}.

Nowadays there is a demand to understand the cooperative behavior 
of relativistic quantum systems. 
It was understood that the vacuum probing 
becomes possible by using high power lasers 
\cite{MTB, MaShu-2006, EPS-Stra, ADiP-2012}. 
With the achievable laser intensity increasing
 at future facilities \cite{ELIILE}, we shall 
encounter novel physical processes such as the radiation reaction dominated regimes 
and then QED processes. 
In a micron focal spot, the laser light 
can generate electron-positron pairs from vacuum 
at intensity two orders of magnitude below \cite{SSB-PRL2010} 
the threshold
corresponding to the QED critical electric field, 10$^{29}$W/cm$^{2}$,
\cite{SAUTER, H-E, SCHWINGER, BLP}.
In this case vacuum begins to act nonlinearly,
its refractive index becomes nonlinearly dependent on the electromagnetic field
strength \cite{NIMA}.
In quantum field theory, particle creation from vacuum has attracted
a great deal of attention, because it provides a typical example of non-perturbative
processes \cite{BLP, DG} in quantum field theory.

Relativistic flying mirrors formed by the laser-driven plasma waves
can be used \cite{MIRR} to achieve an electromagnetic wave intensification
towards reaching the critical field of QED.
Nonlinear QED vacuum properties can be
probed in the future with such strong and powerful electromagnetic pulses.

As regards to other applications, the electromagnetic field intensification and the frequency 
upshift are attractive for research on the development of sources of radiation with 
tunable parameters. The ultimate goal of these studies is the development of a compact 
source of high-intensity X-ray radiation, which is required by a broad range of applications, 
from molecular imaging \cite{15}, which is of high interest for medicine and biology, 
to the diagnostics of thermonuclear plasmas and experiments on laboratory astrophysics \cite{FUNDASTRO, 16, 17, 18} .
We note here the interest towards accelerated relativistic mirrors (see Refs. \cite{QFCS, GRIB, MLobet}) in the context of studying at the terrestrial laboratory conditions 
the Unruh and Hawking effects \cite{PCGM}.

The present review article focuses on the analytical theory 
of nonlinear interactions of intense electromagnetic waves with relativistic 
plasmas related to the RFM concept. 
The experimental results demonstrating the RFM concept 
in the high intensity laser interaction with plasmas
and corresponding computer simulations
can be found in the review articles \cite{MTB, MaShu-2006, MIRR, FUNDASTRO, HOHSD2}
and in the literature cited therein.
}

\section{Relativistic Flying Mirror Concept for Electromagnetic Field Intensification and Frequency Upshifting}

\subsection{Reflection of Electromagnetic Wave from Relativistic Mirror Moving with the Constant Velocity}

\begin{figure}
	    \begin{center}
 \begin{tabular}{c}
    \includegraphics[
	width=16cm,height=6cm
	]{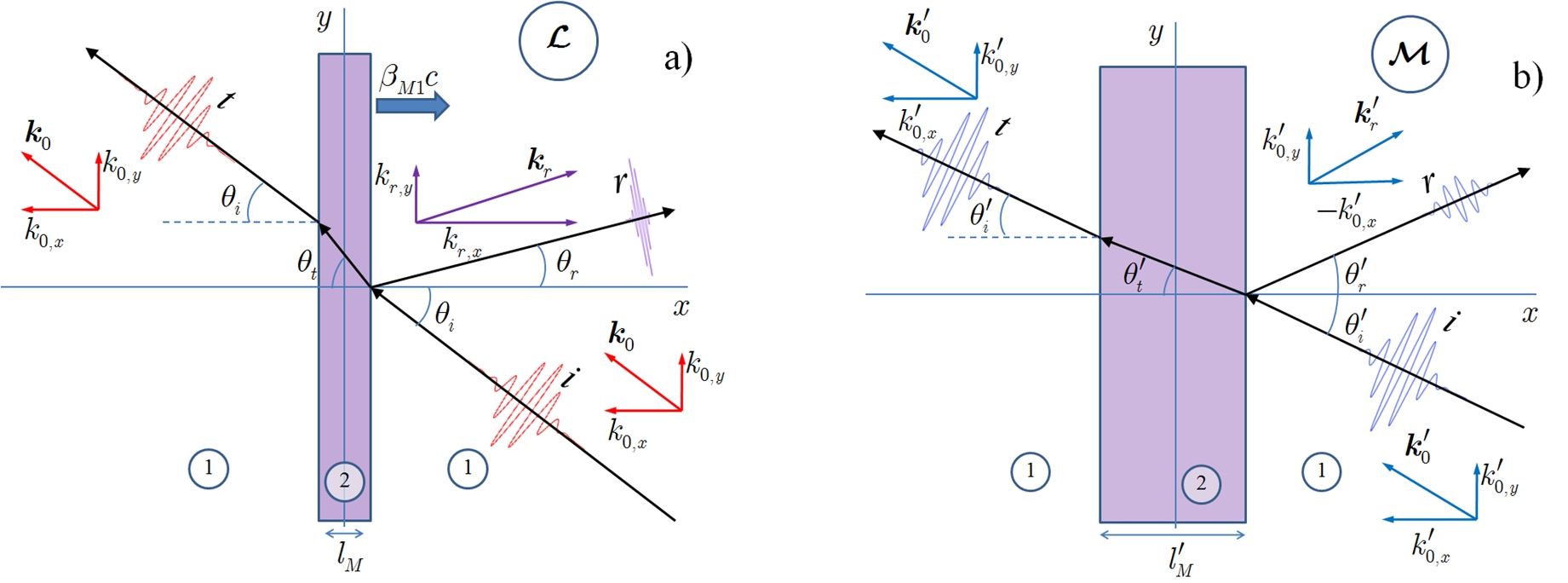}
	\end{tabular}
	\end{center}
	\caption{{ a) Reflection and refraction of the electromagnetic wave at a plasma slab moving with the velocity  $c \beta_M$ along the  $x$-axis.
1 and 2 denote  plasma regions with the density equal to $n_1$ and $n_2$, respectively.
Incident, reflected, and transmitted waves in 
a) the laboratory frame of reference  $\cal{L}$;
b) the mirror rest frame of reference, $\cal{M}$.
}}
		\label{REFLECTION}
	\end{figure}

Change of the frequency and amplitude of an electromagnetic 
wave reflected by moving mirror occurs due to the double Doppler effect.
{
The corresponding theory of the light reflection from the mirror moving in vacuum with arbitrary (subluminal) velocity  
 was formulated by A. Einstein in his paper on the special theory of relativity \cite{Einstein}. 
 It presents a classical example of the application of the Lorentz transformation formalism for solving the problems of classical electrodynamics. 
 The consideration of photon interaction with relativistic mirrors is of great interest to quantum field theory too \cite{QFCS, GRIB}.

The process of the electromagnetic wave reflection from a relativistic mirror 
is characterized by several remarkable features. The reflected 
wave frequency depends on the incidence angle and the mirror velocity. 
We consider reflection and refraction of the 
plane monochromatic wave with the frequency $\omega_0$  and wave vector ${\bf k}_0$
due to the plasma slab with the electron density $n_{e,2}$
moving with the velocity $c \beta_M$  along the  $x$-axis,
as illustrated in Fig. \ref{REFLECTION}.
 
 In the boosted frame of reference, $\cal{M}$, where the mirror is at rest (Fig.  \ref{REFLECTION} b), the wave reflection and refraction are described by Snell's law, 
 i. e. the incidence and reflection angles are equal to each other, $\theta'_i=\theta'_r$, and the refraction angle is given by equation  $\sin \theta^{\prime}_t /\sin \theta^{\prime}_i=n^{\prime}_1/n^{\prime}_2$, 
 where  $n^{\prime}_{1}$  and $n^{\prime}_{2}$  are the refraction indices in media 1 and 2.

 Assuming that  media 1 and 2 are plasma regions with the electron density $n^{\prime}_{e,1}$ and $n^{\prime}_{e,2}$, respectively,  we obtain
 \begin{equation}
\sin \theta^{\prime}_t =\sin \theta^{\prime}_i
\sqrt{
\frac{
1-(\omega_{pe,1}/\omega^{\prime})^2
}{
1-(\omega_{pe,2}/\omega^{\prime})^2
}
},
 \label{eq:sinthetaprime}
 \end{equation}
 where $\omega^{\prime}$ is the wave frequency in the boosted frame of reference and $\omega_{pe,1,2}=\sqrt{4\pi n_{e,(1,2)}e^2/m_e}$. 
 In this frame of reference the wave frequency and the wave number are given by
 \begin{equation}
\omega^{\prime} =\gamma_{M}(\omega_0+k_{0,x}\beta_{M}c),\qquad k_x^{\prime} =\gamma_{M}(k_{0,x}+\omega_0\beta_{M}/c),\qquad k^{\prime}_y=k_{0,y},
 \label{eq:omegaprime1}
 \end{equation}
 with $\gamma_{M}=1/\sqrt{1-\beta_{M}^2}$ being the mirror relativistic Lorentz factor.
As a result of the wave reflection the  $x$ component of the wave vector changes  sign (Fig. \ref{REFLECTION} b). 
 By performing the Lorentz transformation to the laboratory frame of reference, $\cal{L}$, we can find the frequency and wave vector of the reflected wave.  
 Taking into account that  $\sin \theta_i=k_{0,y}/k_0$,  $\sin \theta_r=k_{0,y}/k_r$  and using relationships (\ref{eq:omegaprime1}), we obtain
 }

 \begin{equation}
 \omega_r=\frac{\omega_0 (1+\beta_M^2)+2 \beta_M \sqrt{\omega_0^2-\omega_{pe,1}^2}\cos \theta_i}{1-\beta_M^2}.
 \label{eq:omegar}
 \end{equation}
Here $\omega_0$ and $\omega_r$  are the frequencies of the incident and reflected waves, respectively. The reflection angle
 $\theta_r$ is related to the wave incidence angle  $\theta_i$ as 
 \begin{equation}
 \sin \theta_r=\frac{\omega_0}{\omega_r}\sin \theta_i=
 \frac{\omega_0(1-\beta_M^2)\sin \theta_i}{\omega_0 (1+\beta_M^2)+2 \beta_M \sqrt{\omega_0^2-\omega_{pe,1}^2}\cos \theta_i}.
 \label{eq:thetar}
 \end{equation}
 If the reflection coefficient is equal to unity, as for the ideally reflecting mirror, then during 
 the wave-mirror interaction the ratio of the amplitude of the electric field to its frequency is constant:
 \begin{equation}
 \frac{E_r}{\omega_r}=\frac{E_0}{\omega_0}.
 \label{eq:Er}
 \end{equation}

 Depending on whether the wave and mirror are co-propagating ($\beta_M<0$) or counter-propagating ($\beta_M>0$) 
 in the laboratory frame of reference we have either the frequency downshift or the upshift, 
 the stretching or the compression of the electromagnetic pulse, and decreasing or 
 increasing of the wave amplitude. In the 
 most simple configuration of the wave propagating  in vacuum (i.e. $\omega_{pe,1}=0$)
normally incident on the mirror ($\theta_i=0$), expression (\ref{eq:omegar}) yields
 \begin{equation}
 \omega_r=\omega_0\frac{1+\beta_M}{1-\beta_M}\equiv \omega_0(1+\beta_M)^2\gamma_M^2.
 \label{eq:omegarr}
 \end{equation}

  In the ultrarelativistic limit, when  $\gamma_M\gg 1$, 
 the reflected wave frequency is higher (lower) by a factor $\approx 4 \gamma_M^2$. 
 The reflected wave amplitude changes accordingly.
 
\begin{figure}
	    \begin{center}
 \begin{tabular}{c}
    \includegraphics[
	width=8cm,height=6cm
	]{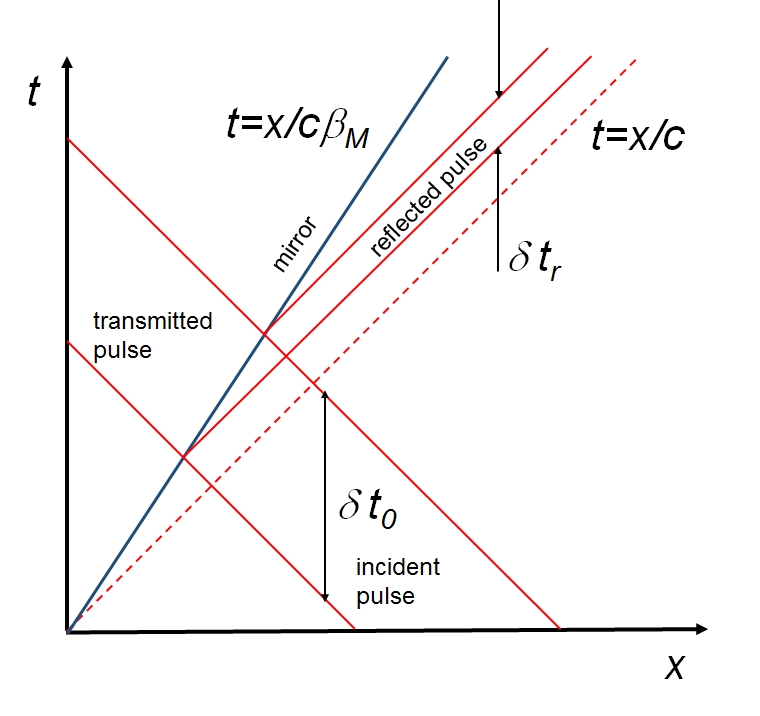}
	\end{tabular}
	\end{center}
	\caption{{ Reflection of an electromagnetic wave at a semi-transparent relativistic mirror. 
	The mirror moves with a constant velocity $c \beta_M$. The world lines corresponding to
the light pulses in vacuum are given by $t=\pm x/c+$constant.} }
		\label{WORLDLINES}
	\end{figure}

 {
 The results obtained can be qualitatively (and quantitatively) explained by considering the light pulse kinematics illustrated by the simple picture in Fig. \ref{WORLDLINES}.
 It shows in the plane $(t,x)$ the light reflection  
 at the relativistic mirror. The mirror position is given by the line $t=x/ \beta_M c$. 
 The world lines corresponding to the incident, reflected and transmitted light pulses 
are straight lines given by $t=\pm x/c+$constant. 
 With the help of elementary geometry we find that the width of the reflected pulse $\delta t_r$ is related to the incident pulse width $\delta t_0$ as 
$\delta t_r=\delta t_i (1-\beta_M)/(1+\beta_M)$ in accordance with expression (\ref{eq:omegarr}).
 }

	\subsection{Light Reflection at the Receding Mirror and Ion acceleration by the Radiation Pressure}
	
{ The light interaction with a co-propagating plasma slab corresponds
to the reflection off a receding mirror.
It results in the frequency down-shift and the energy decrease of the reflected radiation. 
The frequency decreases by a factor  $(1+\beta_M)/(1-\beta_M)$,
Eq. (\ref{eq:omegarr}), which in the limit $\beta_M \to 1$ 
approximately equals $4 \gamma_M^2$. 
The energy of the reflected electromagnetic pulse becomes equal to   ${\cal E}_{las}(1-\beta_M)/(1+\beta_M)$, 
where ${\cal E}_{las}$  is the energy of the  laser pulse incident on the mirror.
  As a result, the energy transferred to the target is given by  $2{\cal E}_{las}\beta_M/(1+\beta_M)$. 
  This process corresponds to the ion acceleration by the radiation pressure of the laser light \cite{VEKSLER, RPDA, SVB-2016}.
 As we see, in the limit $ \gamma_M\gg 1$ almost all the energy of the laser pulse is transferred to the accelerated ions \cite{RPDA}.

In general case, the energy of the  accelerated ions ${\cal E}_{p}$  (here we assume that the ions are protons) 
is determined by the laser energy  ${\cal E}_{las}$ and by the total number of the ions in the region of target irradiated by the laser pulse,  ${\cal N}_{tot}$:
 \begin{equation}
{\cal N}_{tot}{\cal E}_p=2{\cal E}_{las}\frac{\beta_M}{1+\beta_M}.
\label{EpE}
\end{equation}
In the non-relativistic ion energy limit we obtain the scaling:  ${\cal E}_{p}=400\times (10^{11}/{\cal N}_{tot})^2({\cal E}_{las}/10$ J$)^2$MeV.

There are first indications of the ion acceleration by laser radiation pressure obtained in experiments on the laser interaction with thin foil targets \cite{RPDAexp1, RPDAexp2, RPDAexp3}.  

In the case of the incident/reflected light pulse propagating in a 
continuous medium, the light reflection at the moving mirror acquires other properties including the parametric Doppler effect 
and the effects of finite group velocity of the electromagnetic pulse (e.g. see review article \cite{MIRR} and the literature cited therein). This, in particular, becomes important in the case 
of the ion acceleration by the  radiation pressure \cite{RPDA} of ``slow'' light \cite{SLOW, SLOW2, SLOW3}, which may be of special interest in the so called double-sided relativistic mirror configuration \cite{KAGAMI}. 
The changing group velocity medium through which the incident laser pulse propagates can be a preplasma corona formed due to the finite contrast of the laser pulse as discussed in Ref. \cite{CORONA}.
}

If medium, through which the wave propagates, moves one should also take into account 
the Fizeau effect -- the light dragging along by the medium \cite{LL-TP}. 
According to definition of the group velocity of the electromagnetic wave, ${\bf v}_g=\partial_{\bf k}\omega$, 
we have the relationships 
 \begin{equation}
 d\omega={\bf v}_g\cdot d{\bf k}, \quad  d\omega'={\bf v}'_g\cdot d{\bf k}'
 \label{eq:omegaprime}
 \end{equation}
in the laboratory and boosted frames of references. We find
 \begin{equation}
 d\omega=\gamma_M(d\omega'+c\beta_M dk'_x)=\gamma_M({\bf v}'_g\cdot d{\bf k}'+c\beta_M dk'_x)
 \label{eq:domega}
 \end{equation}
and 
 \begin{equation}
\gamma_M\left(1+\frac{\beta_M}{c} v'_{g,x}\right) d\omega
=\gamma_M(v'_{g,x}+c\beta_M) dk'_x+v'_{g,y}dk'_y+v'_{g,z}dk'_z,
 \label{eq:domegapr}
 \end{equation}
which yields 
 \begin{equation}
v_{g,x}=\gamma_M\frac{v'_{g,x}-c\beta_M}{1-\frac{v'_{g,x}\beta_M}{c}}, 
\quad 
v_{g,y}=\gamma_M\frac{v'_{g,y}}{1-\frac{v'_{g,x}\beta_M}{c}},
\quad 
v_{g,z}=\gamma_M\frac{v'_{g,z}}{1-\frac{v'_{g,x}\beta_M}{c}}.
 \label{eq:vgvgprime}
 \end{equation}

 In the case of the wave interacting with the receding mirror, i. e. in the co-propagating 
 configuration, we notice that the wave group velocity should not be lower than the mirror velocity \cite{NNRvg},
 \begin{equation}
v_g=c\frac{\sqrt{\omega_0^2-\omega_{pe}^2}}{\omega_0}\geq c\beta_M.
 \label{eq:vgvm}
 \end{equation}
 This is equivalent to the constraint $\omega_0\geq \omega_{pe} \gamma_M$. {If the mirror moves faster than the laser group velocity, the 
 laser pulse cannot accelerate it. This imposes an upper limit on the energy of accelerated ions \cite{SLOW, SLOW2, SLOW3}.}

 \subsection{Light Reflection at the Accelerated Mirror}
 
\begin{figure}
	    \begin{center}
 \begin{tabular}{c}
    \includegraphics[
	width=15cm,height=10cm
	]{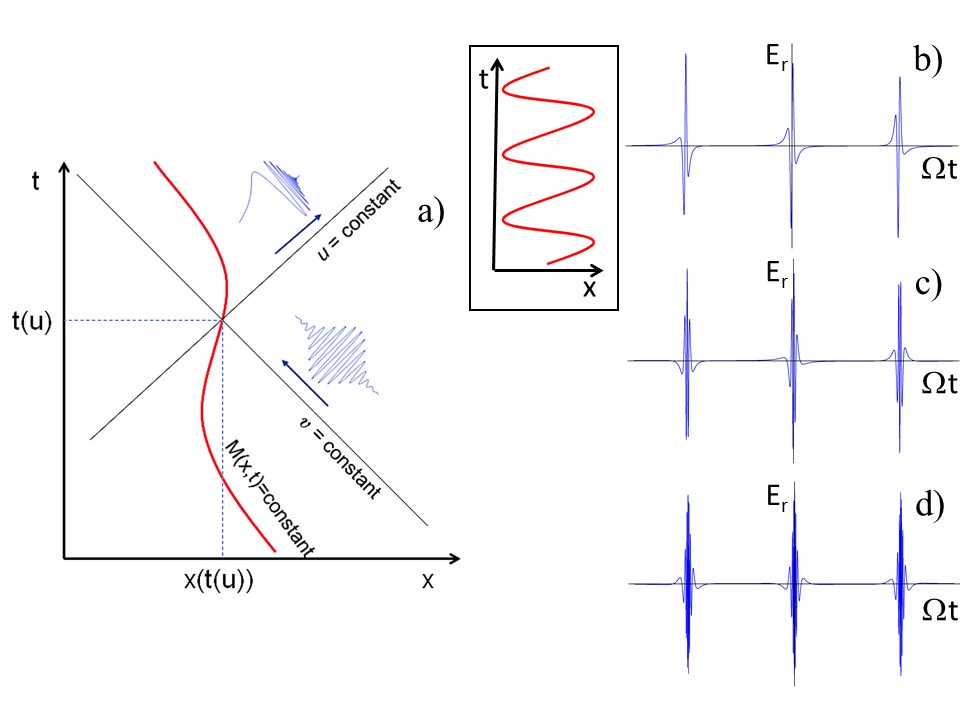}
	\end{tabular}
	\end{center}
	\caption{ 	{
	a) The mirror world line in the  $(t,x)$ plane. The incident, $v=$constant, and reflected,  $u=$constant,  light rays. 
b),c),d) The electric field of the wave reflected from the oscillating mirror for
$a_0=15$ and $\Omega/\omega_0=0.5,1,2$, respectively.
}
	}
		\label{FIGACCMIRR}
	\end{figure}

 {We consider the wave reflection from the accelerated mirror for the case of normal incidence \cite{QFCS, GRIB}.
 The incident and reflected electromagnetic waves in vacuum obey Maxwell's equations from which follows the 
 wave equation for transverse component of the vector potential,
 \begin{equation}
\partial_{tt}A-c^2\partial_{xx}A=0.
 \label{eq:waveEq-xt}
 \end{equation}
The mirror coordinate at time $t$  is determined by equation  ${\cal M}(x,t)=$const, 
so that the mirror is located in the point $x=x_M(t)$, Fig. \ref{FIGACCMIRR} a). 

 We assume that the boundary condition for the field at the mirror is reduced to the Dirichlet condition
 \begin{equation}
\left. A(x,t)\right|_{x=x_M(t)}=0.
 \label{eq:Dirichlet}
 \end{equation}
 This boundary condition is given at the non-stationary surface. 
Following Ref. \cite{GRIB}, 
 we change the independent variables $(t,x)$ to new variables $(\tau,\xi)$, for which the mirror is 
 at the rest and the equation (\ref{eq:waveEq-xt}) remains the wave equation for $A(\xi,\tau)$.
 
 At first, it is convenient to introduce the advanced and retarded coordinates $(u,v)$:
 \begin{equation}
u=t-x/c \quad {\rm and} \quad v=t+x/c.
 \label{eq:uv}
 \end{equation}
In these coordinates Eq. (\ref{eq:waveEq-xt}) takes the form 
 \begin{equation}
\partial_{uv}A=0.
 \label{eq:waveEq-uv}
 \end{equation}
In coordinates $(\bar u, \bar v)$, determined by $u=f(\bar u)$ and $v=g(\bar v)$, equation (\ref{eq:waveEq-uv}) preserves its form except for a factor $f^{\prime}g^{\prime}$:  $\partial_{\bar u\bar v}A=0$.

In new spatial and time coordinates,  $(\xi,\tau)$,  defined as 
 \begin{equation}
\bar u=\tau-\xi /c \quad {\rm and} \quad \bar v=\tau+\xi/c,
 \label{eq:bar uv}
 \end{equation}
equation (\ref{eq:waveEq-xt}) is
 \begin{equation}
\partial_{\tau \tau}A-c^2\partial_{\xi \xi}A=0.
 \label{eq:waveEq-xitau}
 \end{equation}

We choose the functions $f(\tau-\xi /c)$ and $g(\tau+\xi /c)$ to be such that the mirror is at rest at $\xi=0$, 
which implies the equation for the functions $f$ and $g$,
 \begin{equation}
g(\tau)-f(\tau)=2 x_M((g(\tau)+f(\tau))/2),
 \label{eq:fg}
 \end{equation}
transforming the boundary condition (\ref{eq:Dirichlet}) to
 \begin{equation}
\left. A(\xi,\tau)\right|_{\xi=0}=0.
 \label{eq:Dirichlet-xi}
 \end{equation}

We may choose the function $g(\bar v)$ to be equal to $g(\bar v)=\bar v$. This leads to the functional equation for $f$:
 \begin{equation}
\tau-f(\tau)=2 x_M((\tau+f(\tau))/2),
 \label{eq:ftau}
 \end{equation}

 Solution to the boundary problem for Eq. (\ref{eq:waveEq-xitau}) with the boundary condition  (\ref{eq:Dirichlet-xi}) is 
 \begin{equation}
A(\xi,\tau)=A_0\exp\left(\pm i\omega_0 \tau \right)\sin \omega_0\xi.
\label{eq:sol-bound-xitau}
 \end{equation}
In the coordinates $(t,x)$ it can be written as 
 \begin{equation}
A(x,t)=A_0\left[\exp\left(-i\omega_0 v \right)- \exp\left(i\psi_r(u)\right) \right].
 \label{eq:bound-sol-xt}
 \end{equation}
Here variables $u$ and $v$ are defined by Eq. (\ref{eq:uv}) and the function $\psi_r(u)$ is the phase of the reflected wave (the ratio $\psi_r(u)/\omega_0$ is the inverse function to $f(u)$). 
Using Eq. (\ref{eq:ftau}) we obtain that the phase , $\psi_r(u)$,  
is given by 
 \begin{equation}
\psi_r(u)=\int^u\omega_r (u)du=\omega_0 [2 t(u)-u],
\label{eq:psir-u}
 \end{equation}
where we have introduced the frequency of the reflected wave $\omega_r (u)$ related to the phase $\psi_r(u)$.
Here $t(u)$ should be found from the equation
 \begin{equation}
 t(u)=u+ x_M(t(u))/c,
 \end{equation}
in other words, it can be obtained by finding the position of the light ray intersection with the world line of the mirror in the $(t,x)$ plane.}
Here $x_M(t(u))$  is the intersection point coordinate, see Fig. \ref{FIGACCMIRR} a). 

Differentiating the expression for the phase $\psi_r(u)$ with respect to time, we find 
 \begin{equation}
\psi'_r(u)=\omega_0 \frac{1+\beta_M(u)}{1-\beta_M(u)},
 \end{equation}
where  $\beta_M(u)=dx_M(t(u))/cdt$  is the mirror velocity normalized by $c$. 
The derivative of the phase with respect to variable $u$  determined by Eq. (\ref{eq:psir-u}) 
is nothing more than the frequency of the reflected wave, $\omega_r$. 

When the mirror moves with uniform acceleration,  $w_M$, 
a dependence of its coordinate on time is given by the relationship \cite{LL-TP} 
 \begin{equation}
x_M(t)=\frac{c}{w_M}\sqrt{c^2+w_M^2t^2}.
 \end{equation}
In this case the frequency of  light reflected from the uniformly accelerated mirror
in the limit  $t\to \infty$ grows proportionally to the square of the time for positive acceleration, $w_M>0$, 
and decreases inversely proportional to the square of the time for negative acceleration  $w_M<0$: 
\begin{equation}
\omega_r=\omega_0 \frac{\sqrt{c^2+w_M^2t^2}+w_Mt}{\sqrt{c^2+w_M^2t^2}-w_Mt}
\underset{t\rightarrow \infty }{\approx}\left\{ 
\begin{array}{c}
4\omega_0 w_M^2t^2/c^2, \quad w_M>0\\ 
\omega_0 c^2/4w_M^2t^2, \quad w_M<0%
\end{array}%
\right. 
\end{equation}
The amplitude of the reflected wave increases (decreases) in a similar way. 

\section{Thin Electron Layer as a Relativistic Mirror}

The interaction of a sufficiently wide electromagnetic pulse
with a thin foil, in the case where the ponderomotive force of
the pulse significantly exceeds the force caused by the electric
field, which occurs due to electric charge separation, can result
in the formation of a dense electron layer moving in the
direction of electromagnetic wave propagation. The second
counter-propagating electromagnetic pulse is partialy
reflected from a thin electron layer, which, due to the 
above-mentioned double Doppler effect, should lead to the
compression of the reflected pulse and to its frequency
upshift \cite{KuVV-2007}. 
{This scheme has attracted recently significant attention both in theoretical work  \cite{KuVV-PRL, WU, swarm, AAA, MA-2014} and in experiment \cite{KIEFER}.
}

{As a simple model of  the acceleration of a thin electron layer by a
laser pulse, we use the known exact solution of the equations
of motion of a charged particle in the field of a plane
electromagnetic wave \cite{LL-TP}. 
For  the velocity of the electron layer motion along the $x$-axis it yields}  
\begin{equation}
v_{||}=\frac{p_{||}}{m_e \gamma_e}=c\frac{|{\bf a}_{\perp}(u)|^2}{2+|{\bf a}_{\perp}(u)|^2}
\label{eq:vparal}
 \end{equation}
with $u=t-x/c$. The layer coordinate is defined by implicit
equations given in Ref. \cite{LL-TP}. For a
circularly polarized wave, for example, with a constant dimensionless amplitude $a_0$,
we obtain
\begin{equation}
x(t)=\frac{2 x_0+a_0^2 ct}{2+a_0^2},
 \end{equation}
where $x_0$ is the layer coordinate value at the instant of arrival of the
laser pulse, $t_0= x_0/c$.

Using the expression for the electron velocity in Eq. (\ref{eq:vparal}),
we find the corresponding gamma-factor of the relativistic
mirror:
\begin{equation}
\gamma_M(u)=\frac{1}{\sqrt{1-v_{||}^2(u)/c^2}}=\frac{2+|{\bf a}_{\perp}(u)|^2}{2\sqrt{1+|{\bf a}_{\perp}(u)|^2}},
 \end{equation}
It can be seen that in the limit of a large-amplitude
electromagnetic wave, when $|{\bf a}_{\perp}(u)| \gg 1$, the gamma-factor
is proportional to the first power of $|{\bf a}_{\perp}(u)|$.

 \subsection{Light Reflection at the Oscillating Mirror}
 
When an electromagnetic wave reflects from an oscillating mirror, its frequency spectrum 
extends to the high frequency range and the wave breaks up to short wave packets, as shown in Fig. \ref{FIGACCMIRR} b), c), and d). 
According to Eq. (\ref{eq:omegar}) the wave frequency increases 
by a factor approximately equal to $4 \gamma_M^2$. 
By the same factor the maximal electric field increases according to Eq. (\ref{eq:Er}). 

{The relativistic oscillation mirror (ROM) concept has been proposed in Ref. \cite{ROM} 
as a mechanism of high order harmonic generation when an overdense plasma 
is irradiated by relativistically intense laser radiation. 
High frequency radiation generation
has been demonstrated in experiments with multi-terawatt lasers, see review article \cite{HOHSD2}, literature cited therein and discussions of alternative mechanisms}.

As an illustration of this process we consider a thin electron layer oscillating under the action 
of a linearly polarized electromagnetic wave of the frequency  $\Omega$. 
The wave's electric field is parallel to the $y$  axis: $E_y=E_0 \cos(\Omega v)$ with $v=t+x/c$.
The second electromagnetic wave with the amplitude $a_i$  is reflected from the electron layer 
as from the mirror. 
In order to describe the electron layer motion we use here the results of an exact solution 
of the problem on the electric charge dynamics in the field of an electromagnetic wave \cite{LL-TP}, 
which yield the parametric dependences of the charge coordinates and momentum components on time in the frame of
reference where the charge is on average at the rest: 
\begin{equation}
x=\frac{a_0^2}{4(2+a_0^2)}\sin 2\eta, \quad y=\frac{a_0}{\sqrt{1+a_0^2/2}}\cos \eta, \quad z=0,
\label{eq:xyosc}
 \end{equation}
\begin{equation}
t=\eta+\frac{a_0^2}{4(2+a_0^2)}\sin 2\eta, 
\label{eq:t-etaosc}
 \end{equation}
\begin{equation}
p_x=\frac{a_0^2}{4\sqrt{1+a_0^2/2}}\cos 2\eta, \quad p_y=a_0 \sin \eta, \quad p_z=0.
\label{eq:pxosc}
 \end{equation}
Here the coordinates and time are normalized by $c/\Omega$  and  $\Omega^{-1}$,  respectively,
$\eta=\Omega v$, 
the momentum is measured in units of $m_e c$, the normalized amplitude of the wave equals  $a_0=eE_0/m_e \Omega c$. 
{The incident wave has a constant amplitude.}

Now we can calculate the phase and frequency of the reflected wave and find the electric field 
\begin{equation}
E_r(t)=E_i\omega^{-1}_0\psi'_r(t) \cos \psi_r(t).
 \end{equation}
{This is shown in Figs. \ref{FIGACCMIRR} b), c), and d) for $\Omega=\omega_0/2$, $\Omega=\omega_0$, and $\Omega=2 \omega_0$,
 respectively, where the mirror oscillation amplitude is characterized by $a_0=15$.} 
The reflected radiation has the form of a sequence of wave packets with the frequency in the limit
 $a_0 \gg 1$   approximately equal to  $\omega_0 a_0^2/2$ and with the wave packet duration 
 $\approx 2\pi/\Omega a_0$. 
 Here we assume that the second wave propagates in the direction 
 opposite to that of the wave which drives the mirror oscillations. 
If the second wave and the wave driving the mirror oscillations propagate 
in the same direction then the electric field in the reflected wave has 
a form of a sequence of short pulses with the width  $\approx \pi/\Omega a_0$ with no high frequency filling. 
{
We note that the generation of ultrashort electromagnetic pulses observed 
in computer simulations of the relativistically strong electromagnetic wave interaction 
with an overdense plasma target in Ref. \cite{HOHSD} has been explained by the authors within the framework of the ROM concept.}

\begin{figure}
	    \begin{center}
 \begin{tabular}{c}
    \includegraphics[
	width=10cm,height=6cm
	]{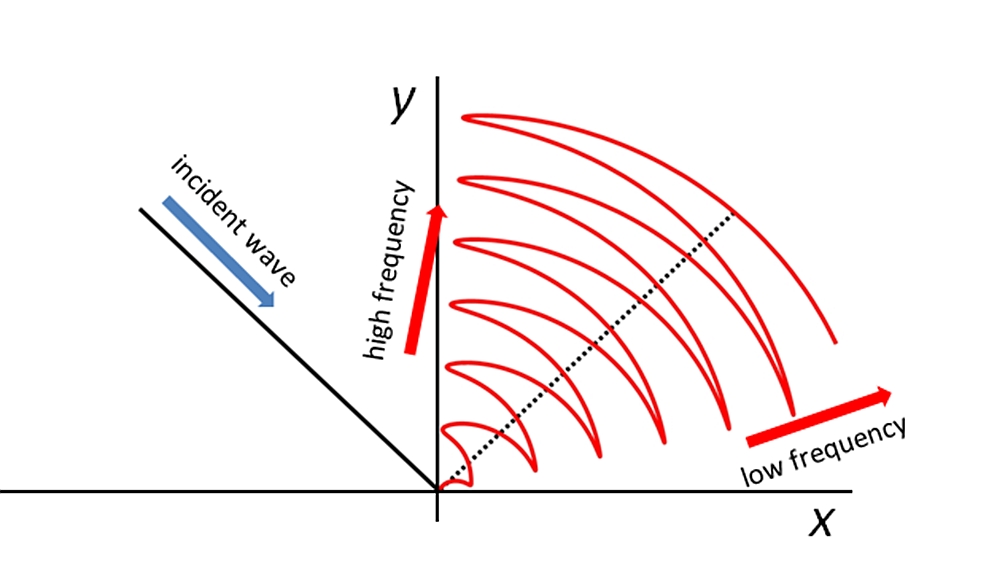}
	\end{tabular}
	\end{center}
	\caption{{ The instantaneous distribution of photons  
reflected by an oscillating mirror in the $(x,y)$ plane, for an obliquely incident wave. The photons with the upshifted frequency propagate closer to the 
mirror normal direction, while low frequency photons are emitted at a grazing angle with respect to the mirror plane. }}
		\label{AngSec}
	\end{figure}

According to Eq. (\ref{eq:thetar}) the angle of light reflection $\theta_r$ depends on the mirror velocity.
The larger the mirror positive velocity the closer the reflected light propagation direction is to the mirror normal. 
For photons reflected in the mirror negative velocity phase, the propagation direction 
is close to the mirror surface. The photons with the upshifted frequency are localized in the region close to the 
mirror normal direction. {The low frequency photons propagate at a grazing angle with respect to the mirror plane}.
Using expressions (\ref{eq:xyosc}, \ref{eq:t-etaosc}, \ref{eq:pxosc}) 
we can find the instantaneous photon distribution in the $(x,y)$ plane for the electromagnetic wave 
reflected by oscillating mirror which is shown in Fig. \ref{AngSec}.
{This can be considered as a lighthouse effect. The attosecond lighthouses from oscillating plasma mirrors have been demonstrated in Ref. \cite{WHEELER}, 
where an isolated nonlinearly generated attosecond pulse train has been obtained by rotating the instantaneous wavefront direction of an intense few-cycle laser field. 
Here we show that the angle separation of high and low frequency photons can also occur due to the relativistic dependence of the reflection angle on the instantaneous velocity of the oscillating mirror.}

 \subsection{Reflection Coefficient of Electromagnetic Radiation from a Thin Electron Layer}
 
Here we discuss the reflection of electromagnetic waves from a thin electron layer. 
In order to calculate the coefficients of reflection, ${\rm {\bf R}}$,  and transmission, ${\rm {\bf T}}$,   
we consider the interaction of electromagnetic waves with a strongly nonuniform electron distribution. 
The electrons reside in a layer with the density $n$, which determines the plasma frequency as a function of coordinate and time, 
\begin{equation}
\omega_{pe}=\sqrt{\frac{4\pi n(x-c\beta_Mt)e^2}{m_e \gamma_M}}.
 \end{equation}

	Performing the Lorentz transformation to a boosted frame reference moving 
	with the velocity $c\beta_M$, we find that in this frame of reference, 
	the wave equation describing the electromagnetic 
wave interaction with the plasma slab takes the form
\begin{equation}
\frac{d^2 a(\zeta)}{d\zeta^2}+q^2(\zeta)a(\zeta)=0,
\label{eq:wavezeta}
 \end{equation}
where 
\begin{equation}
q^2(\zeta)=s^2+\nu_e(\zeta) .
\label{eq:qzeta}
 \end{equation}
{Here $\nu_e(\zeta)=\omega_{pe}^2/c^2$,
and variables
\begin{equation}s^2=(\omega'/c)^2-k_y^2, \quad \zeta=\gamma_M(x-c\beta_M t), \quad  t',  \quad k' \quad {\rm and}   \quad \omega'
 \end{equation}
are the square of the wave number,
coordinate, time,  incident wave number and frequency in the boosted frame of reference, respectively}.
The vector potential is normalized by  $m_ec^2/e$, 
and its dependence on time  $t'$ and coordinates  $\zeta$ and  $y$ is 
\begin{equation}
a(\zeta)=\frac{eA_z(\zeta)}{m_ec^2}\exp\left[-i\left(\omega't'-k_yy\right)\right].
 \end{equation}
The reflection coefficient of electromagnetic radiation from a thin electron 
layer we find in a similar way as is done in the classical problem of scattering. 
Assuming that the layer is in the rest frame of reference, 
we take the distribution of the electron density in the form proportional 
to the Dirac delta function: $n(\zeta)=n_0l_0\delta(\zeta)$,
where $l_0$  is the electron layer thickness.
 
In this case the function $\nu_e(\zeta)$ in Eq. (\ref{eq:qzeta}) is $\nu_e=g_{\delta}\delta(\zeta)$ 
with $g_{\delta}=k^2_pl_0$ and $k_p=\omega_{pe}/c$.
Integrating this equation over $\zeta$  in the interval $-\varepsilon<\zeta<\varepsilon$  
and letting $\varepsilon$  go to zero, 
we obtain the condition for the jump of the derivative $da/d\zeta$  on the boundary $\zeta=0$: 
\begin{equation}
\left.\frac{da}{d\zeta}\right|_{+0}-\left.\frac{da}{d\zeta}\right|_{-0}=g_{\delta}a(0).
\label{eq:boundzeta0}
 \end{equation}
The function  $a(\zeta)$ is continuous in the point  $\zeta=0$. 
The solution to Eq. (\ref{eq:wavezeta}), describing the wave reflection 
from a thin layer can be represented as
\begin{equation}
a(\zeta)=\left\{ 
\begin{array}{c}
\exp(is\zeta)+\rho  \exp(-is\zeta),\quad \zeta \geq 0,\\ 
\tau \exp(is\zeta),\quad \quad\quad \quad \quad \quad \quad \zeta \leq 0,
\end{array}%
\right. 
 \end{equation}
where $\rho$  and $\tau$  are related to each other by the equations following 
from the boundary condition (\ref{eq:boundzeta0}): 
\begin{equation}
1+\rho(s)=\tau(s),
 \end{equation}
\begin{equation}
iq[1-\rho(s)-\tau(s)]=g_{\delta}\tau(s).
 \end{equation}
As a result the reflected wave amplitude is equal to 
\begin{equation}
\rho(s)=-\frac{g_{\delta}}{2is+g_{\delta}}.
 \end{equation}
From this expression we obtain the reflection coefficient for the number of photons 
\begin{equation}
{\rm {\bf R}}_{\delta}=\frac{g_{\delta}^2}{4s^2+g_{\delta}^2}.
\label{eq:R-delta}
 \end{equation}
For large values of the electron surface density, $n_0l_0$, i. e. for  $g_{\delta}^2 \gg 4s^2$, 
the reflection coefficient is close to unity. In the opposite limit, we can neglect $g_{\delta}^2$  
as compared with $4s^2$  in the denominator of the right-hand side 
of the expression (\ref{eq:R-delta}) and obtain 
{
\begin{equation}
{\rm {\bf R}}_{\delta}\approx\frac{g_{\delta}^2}{4s^2}\approx \frac{\varepsilon_e^2}{\gamma_M^2}.
\label{eq:R-delta1}
 \end{equation}
 where the dimensionless parameter introduced in Ref. \cite{FOIL}
\begin{equation}
\varepsilon_e =n_0l_0\lambda_0r_e=\frac{2\pi n e^2 l}{m_e \omega_0 c}
\label{eq:eps_e}
 \end{equation}
 characterizes the relativistic transparency of a thin foil target.}
It is seen that the reflection coefficient is proportional to the square 
of the electron density in the surface layer, $n_0 l_0$, i. e. the reflection is coherent.

{In the thin electron layer paradigm of a relativistic mirror,
 the effects of Coulomb explosion and transverse modulations 
caused by the electromagnetic pulse non-uniformity can deteriorate the property of the reflected radiation. 
As shown in Ref. \cite{swarm}, sufficiently homogeneous laser pulses 
are required to form thin electron layers continuous in the transverse direction. 
When this condition is not fulfilled in the interaction of laser radiation 
with a thin foil, a cloud of ultrashort bunched electrons  (a swarm of electron bunches) is formed, resulting in loss of coherency and 
broadening of the reflected pulse frequency spectrum. Contrary to that the relativistic flying mirror concept using the wake plasma wave (considered below) 
shows more stable and robust behavior. }

\subsection{Relativistic Transparency of a Thin Plasma Layer}

The 1D electrodynamics model, {where the role of a point charge
is played by an infinitely thin foil} \cite{FOIL, ThinFoil1, Brat},
has been extensively used in studying
the problem of relativistic thin plasma layer transparency,
particularly for the purposes of the laser pulse shaping \cite{FOIL} (see
also the experimental paper \cite{REED}), in high order harmonics
generation \cite{SLIDING1, SLIDING2, ThinFoil2}, in laser ion acceleration \cite{ThinFoil2}, 
in the analysis of the radiation friction effects \cite{RAD4}
and in the
generation of coherent extremely high intensity x-ray pulses
by relativistic mirrors \cite{BET-03, KuVV-2007}.

Using the results of Refs. \cite{FOIL, ThinFoil1,ThinFoil2}, we consider the case
of normal incidence of a plane electromagnetic wave on an
infinitely thin foil. The foil is located in the plane $x = 0$. The
interaction of the wave with the foil is described by Maxwell's
equations for the vector potential $A(x,t)$ which yield
\begin{equation}
\partial_{tt}{\bf A}-c^2 \partial_{xx}{\bf A}=4\pi c \delta(x) {\bf J}({\bf A}).
\label{eq:A-foil}
 \end{equation}
The term on the r.h.s. describes the electric current in the
foil and the delta-function represents its sharp localization.
The solution to the initial problem for Eq. (\ref{eq:A-foil}) yields
\begin{equation}
{\bf A}(x,t)={\bf A}_0(x,t)+2\pi \int_0^{t-|x|/c} {\bf J}({\bf A}(0,\tau))d\tau,
\label{eq:A-foilsol}
 \end{equation}
where ${\bf A}_0(x,t)$ is a known function describing the incident 
electromagnetic wave.
Assuming $x = 0$ on both sides of Eq. (\ref{eq:A-foilsol}) and taking the
derivative with respect to time, we obtain
\begin{equation}
\frac{d {\bf A}(0,t)}{dt }=\frac{d {\bf A}_0(0,t)}{dt }+2\pi {\bf J}({\bf A}(0,t).
\label{eq:A-foilsol1}
 \end{equation}
In this way a nonlinear boundary problem for a system
of partial differential equations is reduced to the ordinary
differential equation for the field inside the foil, ${\bf A}(0,t)$.

In the case of the circularly polarized electromagnetic wave, it is convenient to write the longitudinal 
and transverse components of electron momentum,  $p_{x}$ and $p_y {\bf e}_y+p_z {\bf e}_z$,  
as a combination of vectors, having components rotating with angular frequency  $\omega$, 
\begin{equation}
\left( 
\begin{array}{c}
p_{1} \\ 
p_{2} \\ 
p_{3}%
\end{array}%
\right) =\left( 
\begin{array}{ccc}
1 & 0 & 0 \\ 
0 & \sin \omega t & -\cos \omega t \\ 
0 & \cos \omega t & \sin \omega t
\end{array}%
\right) \left( 
\begin{array}{c}
p_{x} \\ 
p_{y} \\ 
p_{z}%
\end{array}%
\right) 
\label{eq:p123rot}
 \end{equation} 
where $p_2$  and  $p_3$ are the components parallel and perpendicular to the electric field. 
Taking into account the generalized electron momentum
conservation, ${\bf p}- e{\bf A}/c =$constant, and the relationship between
the electric current and the electron velocity, ${\bf J} =
-en_el{\bf v} = -en_elc{\bf p}/(m_ec^2 + p^2)^{1/2}$, where $n_e$ and $l$ are the
electron density and the foil thickness, respectively, we find that the 1D
equation (\ref{eq:A-foilsol1}) for the stationary motion of a thin foil interacting
 with a rotating electric field can be written in
the form 
\begin{equation}
p_2=\varepsilon_{e}\frac{p_3}{\gamma_e},
\label{eq:p2-foilsol}
 \end{equation}
\begin{equation}
p_3=a-\varepsilon_{e}\frac{p_2}{\gamma_e}
\label{eq:p3-foilsol}
 \end{equation}
 where $\gamma_e=\sqrt{1+p_2^2+p_3^2}$ and {the parameter $\varepsilon_{e}$ is defined by Eq. (\ref{eq:eps_e})}.
 
 Solving this system of algebraic equations we obtain 
\begin{equation}
p_2=\frac{\varepsilon_{e}}{\sqrt{2}a}\frac{\sqrt{(1-a^2+\varepsilon_{e}^2)^2+4a^2}-(1-a^2+\varepsilon_{e}^2)}
{\sqrt{(1-a^2+\varepsilon_{e}^2)^2+4a^2}+(1-a^2+\varepsilon_{e}^2)},
\label{eq:p2-foilsolv}
 \end{equation}
\begin{equation}
p_3=\frac{\sqrt{(1-a^2+\varepsilon_{e}^2)^2+4a^2}-(1-a^2+\varepsilon_{e}^2)}
{2 a}.
\label{eq:p3-foilsolv}
 \end{equation}

 In the limit of a relatively weak electromagnetic field when $1 \ll a \ll \varepsilon_{e}$  
 expressions (\ref{eq:p2-foilsolv}) and (\ref{eq:p3-foilsolv}) have the asymptotics
\begin{equation}
p_2=\frac{\varepsilon_{e}}{1+\varepsilon_{e}^2}a-\frac{\varepsilon_{e}^2}{(1+\varepsilon_{e}^2)^3}a^3+O(a^5),
\label{eq:p2-foilasym}
 \end{equation}
\begin{equation}
p_3=\frac{a}{1+\varepsilon_{e}^2}-
\frac{\varepsilon_{e}(1-\varepsilon_{e}^2)}{(1+\varepsilon_{e}^2)^3}a^3+O(a^5).
\label{eq:p3-foilasym}
 \end{equation}
 In the opposite limit for  $1 \ll \varepsilon_{e}^{-1/3} \ll a$ the asymptotics are
\begin{equation}
p_2=\varepsilon_{e}-\frac{\varepsilon_{e}(1+\varepsilon_{e}^2)}{2 a}+O(a^{-3}),
\label{eq:p2-foilasyl}
 \end{equation}
\begin{equation}
p_3=a-
\frac{\varepsilon_{e}^2}{a}+O(a^{-3}).
\label{eq:p3-foilasyl}
 \end{equation}

 As we can see, in the  range of parameters, which corresponds to a relatively low electromagnetic 
 field amplitude  $a \ll \varepsilon_{e}$, the component of the electron momentum $p_3$  parallel 
 to the electric field is much larger than the perpendicular component  $p_2$. 
 In the limit of a strong electric field $a \gg \varepsilon_{e}$ we have $p_2 \gg p_3$,
 that is, the electron momentum is almost perpendicular to the instantaneous direction of the electric field. 
 
 Using obtained expressions (\ref{eq:p2-foilsolv}) and (\ref{eq:p3-foilsolv}) 
 and equation (\ref{eq:A-foilsol}) we can find 
 the amplitudes of the reflected and transmitted waves. 
 If the incident wave has moderate intensity,  $a \ll \varepsilon_{e}$, 
 amplitudes of the reflected and transmitted waves are 
\begin{equation}
a_r=\frac{a \varepsilon_{e}}{\sqrt{1+\varepsilon_{e}^2}} 
\quad {\rm and} \quad a_t=\frac{a }{\sqrt{1+\varepsilon_{e}^2}}.
\label{eq:p3-foilarat}
 \end{equation}
 We see that for $\varepsilon_{e} \gg 1$  the wave is reflected almost totally, 
 while in the limit $\varepsilon_{e} \ll 1$  the electron layer is transparent.
 
For ultrahigh intensity,  $a \gg 1$, if  $a \gg \varepsilon_{e} \gg 1$, the amplitude of the reflected 
wave equals  $\varepsilon_{e}$, while the change in the transmitted wave amplitude,  $\sqrt{a^2-\varepsilon_{e}^2}$, 
is small. This regime can be explained by taking into account the fact that the electric current 
density in a plasma cannot exceed  $j_{lim}=enc$ since the electron velocity is limited 
by the speed of light in vacuum.

We see that the condition for the foil to be transparent to the electromagnetic radiation 
in the limit of moderate intensity $a \ll 1$ is  $\varepsilon_{e} \ll 1$. 
It can be rewritten as  $\omega \gg \omega_{pe}(\pi l/\lambda_p)$, where  $\lambda_p=2\pi c/\omega_{pe}$, 
which differs from the transparency condition for uniform plasmas by the factor  $\pi l/\lambda_p$.

For relativistically strong waves with   $a \gg 1$, a foil with $\varepsilon_{e} \gg 1$ 
is transparent if  $a \gg \varepsilon_{e}$. 
This condition can be cast in the form
\begin{equation}
\omega \gg \frac{\omega_{pe}^2l}{2c a},
\label{eq:p3-foiltrans}
 \end{equation}
while the transparency condition for a uniform extended plasma
irradiated by relativistically strong radiation can written as
\begin{equation}
\omega \gg \frac{\omega_{pe}}{(1+a^2)^{1/4}}.
\label{eq:p3-platrans}
 \end{equation}

{
The expressions for the reflected and transmitted waves have been obtained in the frame of reference,
 where the plasma layer is at rest. The parameters of the reflected and transmitted waves in the boosted frame of reference, 
 i.e. the electric field amplitudes, $E_r$, $E_t$, and the frequencies, $\omega_r$, $\omega_t$, can be found using Lorentz transformations while
taking into account the 
 Lorentz invariance of the normalized amplitudes $a_r\propto E_r/\omega_r$ and $a_t\propto E_t/\omega_t$.
}
 
\section{Nonlinear plasma waves as relativistic mirrors}

An approach to generate high frequency radiation based on the
concept of the relativistic flying mirror
utilizes a plasma shell traveling close to the speed of light 
and acting as a relativistic mirror. 
Reflected light undergoes
double Doppler frequency up-shift, compression, intensification
due to relativistic effects
and focusing (if the plasma shell is curved).

Relativistic plasma oscillation waves can be generated in the plasma in various ways. 
When the laser radiation interacts with an inhomogeneous plasma, e.g. with the plasma 
corona formed at the front surface of a solid target, Langmuir oscillations 
are excited in the vicinity of the plasma critical density surface, 
which is also called the plasma resonance region. {These Langmuir oscillations 
take the form of a localized wave packet whose group velocity is zero
while  its phase velocity is
 directed against the gradient of the plasma density \cite{VLG, BKS}.  Sufficiently high amplitude oscillations eventually break 
 producing thin dense electron layers, which can play a role of relativistic mirrors for the incident electromagnetic radiation. 
 We note that thin electron layers are also formed during the so-called ``vacuum heating'' \cite{VacHeat} at the plasma vacuum interface. 
Scattering of the electromagnetic wave at such thin electron layers have been associated  in Ref. \cite{ROM} (see Fig. 7 (d) therein) with  
high order harmonic generation (see also the recent work \cite{COUSENS}). }

Another, one of the most widely used methods, involves the use of a relatively short 
high-intensity laser pulse with a length less than or of the order of the plasma wavelength 
whose normalized amplitude is large enough. Such a laser pulse, propagating in a collisionless plasma, 
excites the plasma waves in a wake behind it \cite{T-D, ECL}. {Nonlinear wake waves can break forming thin dense electron layers (relativistic plasma mirrors) at the wave crest.}
Various schemes {of the relativistic mirrors} were 
described \cite{BET-03, MIRR, SSB-SphW, MLobet}
and experimentally demonstrated \cite{K1, P1, K2} proving a feasibility of 
the RFM concept.

\subsection{Plasma Oscillations Excited in Near-Critical Inhomogeneous Plasma}

Let us consider an electromagnetic wave obliquely incident on an inhomogeneous 
plasma forming an angle $\theta_i$  with respect to the unperturbed plasma surface. 
{Unperturbed electron density in the vicinity of the 
critical density surface (resonance surface)
 can be approximated by a linear function: $n_0(x)=n_0 (1-x/L)$.
Here $L$ is the scale length of the plasma inhomogeneity.
Due to the resonance condition we have $n_0=m_e \omega_0^2/4\pi e^2$, i.e. the Langmuir frequency locally depends on the coordinate as $\omega_{pe}(x)=\omega_0 \sqrt{1-x/L}$.}
In the vicinity of the plasma resonance surface, the field of 
the p-polarized wave with the electric field in the plane 
of incidence has a singularity \cite{VLG}. 
At the singularity the electric field component parallel to the gradient of the plasma density is given by 
\begin{equation}
E_x=\frac{E_d\sin \theta_i}{\varepsilon(x)},
\label{eq:ExPlaRes}
 \end{equation}
where $E_d$  is the driver field related to the incident wave amplitude as 
\begin{equation}
E_d=\frac{E_0}{\sqrt{2\pi \omega_0 L/c}}\phi\left(\left(\frac{\omega_0 L}{c}\right)^{1/3}\sin \theta_i\right)
\label{eq:EdPlaRes}
 \end{equation}
with the maximum $\phi(0.8)\approx 1.2$, and the dielectric constant 
\begin{equation}
\varepsilon(x)=1-\frac{\omega^2_{pe}(x)}{\omega_0(\omega_0+{i}\nu_{eff})}\underset{x\rightarrow 0 }{\approx}
\frac{x}{L}+{i}s.
\label{eq:EpsPlaRes}
 \end{equation}
 Here the parameter  $s=\nu_{eff}/\omega_0$ with the effective decay rate $\nu_{eff}$ describes the effects of small dissipation, 
 weak nonlinearity, dispersion, etc. \cite{BKS}. 
 
 The maximum of the $x$-component of the electric field in the plasma resonance region is about  $E_m=E_d/s$. 
 The width of the plasma resonance region is equal to  $\Delta x=sL$.
 
 For the sake of simplicity we consider here small amplitude plasma oscillations. 
 For the analysis of nonlinear plasma oscillations in the resonance region see Refs. \cite{BKS} and \cite{ROM}. 
{ The motion of the electron fluid element in the $x$ direction is described by the equation 
\begin{equation}
\ddot \xi_x=-e E_x/m_e, 
\end{equation}
 where $ \xi_x$ is the $x$-component of the electron displacement.
From this equation, using expressions (\ref{eq:ExPlaRes}) and (\ref{eq:EpsPlaRes}) we obtain for $ \xi_x$
\begin{equation}
\xi_x=r_E\frac{s \sin(\omega_0t)-(x/L)\cos(\omega_0t)}{(x/L)^2+s^2}
\label{eq:KsixPlaRes}
 \end{equation}
 with  $r_E=eE_d \sin\theta_i/m_e\omega_0^2$ being the amplitude of the electron quiver motion.
 As we see Eq. (\ref{eq:KsixPlaRes}) describes the localized Langmuir wave packet of width $\Delta x=s L$ with the plasma oscillations inside having the form of a wave propagating against 
 the density gradient with the phase velocity equal to  
\begin{equation}
v_{ph}=\omega_0 \Delta x=s \omega_0 L. 
\label{eq:vphPlaRes}
 \end{equation}
 }
 
\begin{figure}
	    \begin{center}
 \begin{tabular}{c}
    \includegraphics[
	width=16cm,height=8cm
	]{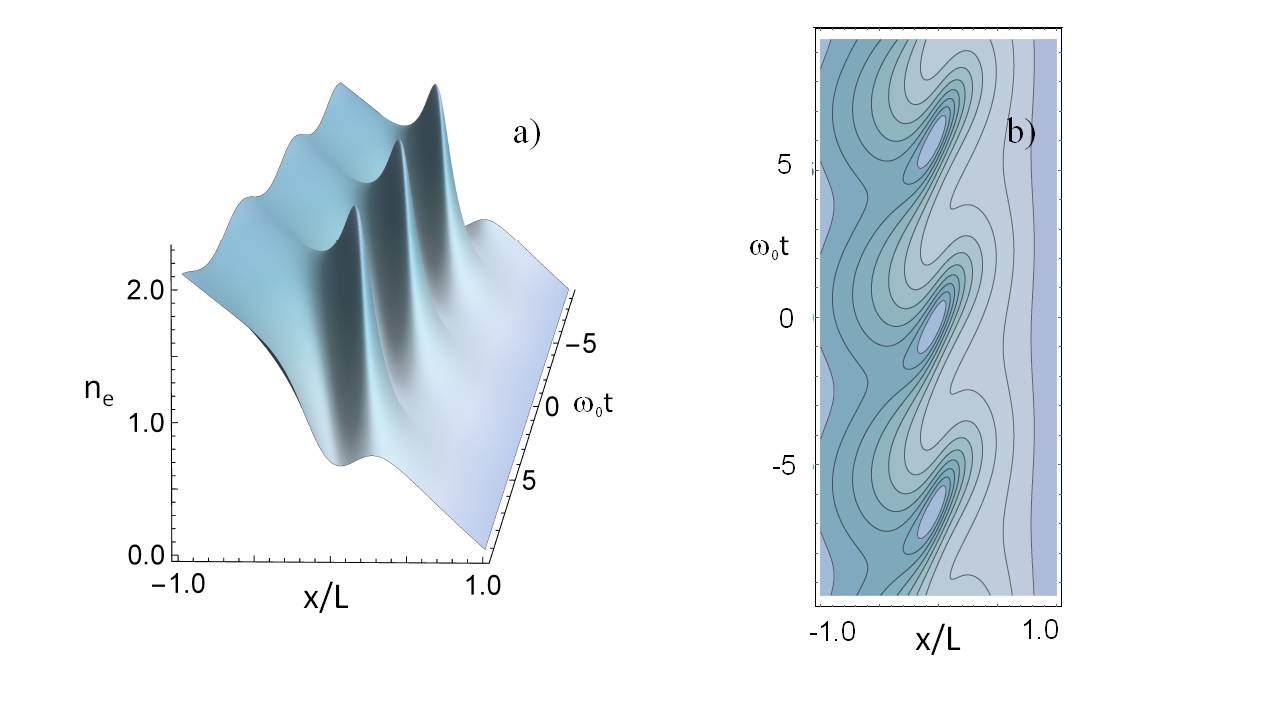}
	\end{tabular}
	\end{center}
	\caption{ {The electron density modulations in the vicinity of the plasma resonance region given by Eqs.  (\ref{eq:KsixPlaRes}) and (\ref{eq:densPlaRes}) for $r_E=0.05$, $s=0.3$ and $L=1$.  
	a) $n_e$ vs $x$ and $t$.
b) Constant density contours in the $(x,t)$ plane.}}
		\label{FIGPLASRES}
	\end{figure}

\begin{figure}
	    \begin{center}
 \begin{tabular}{c}
    \includegraphics[
	width=8cm,height=8cm
	]{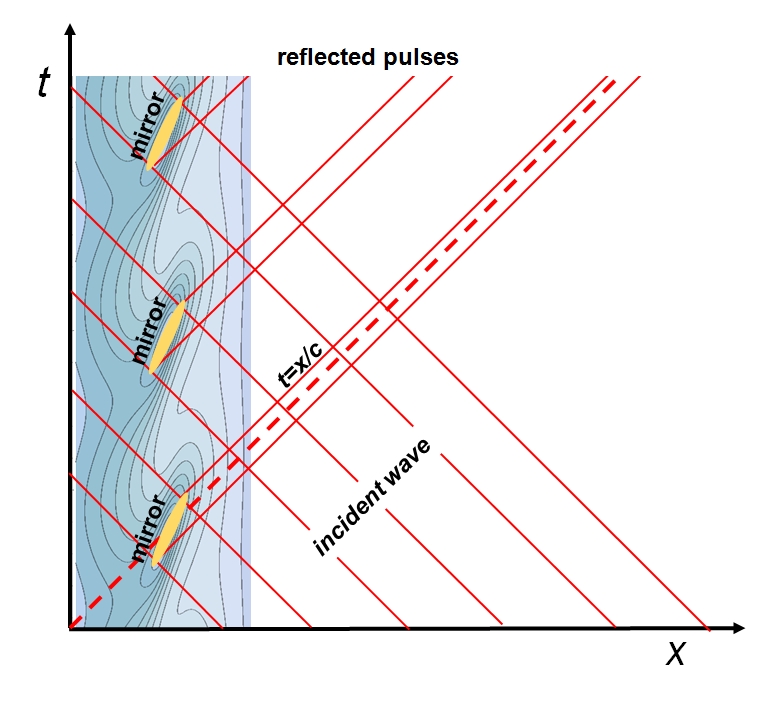}
	\end{tabular}
	\end{center}
	\caption{{ Generation of  ultra-short pulses in the wave reflected at periodically appearing and disappearing mirrors.}}
	\label{ATTO-PULSES}
	\end{figure}

 This causes the electron density modulations. {Using the known dependence of the function $\xi_x$ on time and coordinate we can calculate the electron density. It is given by
\begin{equation}
n_{e}\approx n_0 (x)(1-\partial_x \xi_x). 
\label{eq:densPlaRes}
 \end{equation}
 According to Eqs. (\ref{eq:KsixPlaRes}) and (\ref{eq:densPlaRes})  the density modulations have the form of spikes }
 appearing and disappearing in the plasma resonance region as shown in Fig. \ref{FIGPLASRES}. 
 In other words, we have plasma mirrors periodically appearing and disappearing 
 at the plasma critical surface. {The electromagnetic wave reflection at 
such mirrors results in the generation 
 of  a train of ultra-short pulses as illustrated in Fig. \ref{ATTO-PULSES}.}
Each pulse duration is about $\Delta T=g^{-1}(2\pi/\omega_0)$.
 The frequency upshifting factor $g=\omega_r/\omega_0$ 
 according to Eq. (\ref{eq:omegarr}) is  $g=\gamma_{ph}^2(1+\beta^2_{ph})$. Here $\beta_{ph}=v_{ph}/c$ 
 and $\gamma_{ph}=1/\sqrt{1-\beta_{ph}^2}$.

 Below we shall see that 
 the electromagnetic wave reflection is maximal for the breaking plasma oscillations. 
 At the wave breaking threshold the electron velocity equals the phase velocity of the wave, i.e.  
 $\gamma_e=\gamma_{ph}$. 
 In the case of plasma oscillations driven by the relativistically intense electromagnetic 
 wave with the normalized amplitude  $a_0$ the electron 
 quiver energy Lorentz factor approximately equals  $\gamma_e=a_0$. 
 As a result we have for the frequency upshifting factor  $g\approx 2 a_0^2$.
 
{We note that the train of almost identical short pulses, produced via 
reflection off
the periodically appearing and disappearing relativistic mirrors,
has a spectrum consisting of high-order harmonics.
The width of an individual harmonic is inversely proportional to the 
number of
short pulses, which is determined by the duration of the incident 
electromagnetic wave
and by the life-time of the plasma resonance region.}

	\subsection{Nonlinear Wake Wave Excited by a Short Laser Pulse in Underdense Plasmas}
	
	Finite amplitude plasma oscillations excited in an underdense plasma by a short relativistically 
	intense laser pulse can be described using the system of Maxwell's equations and 
	the cold hydrodynamics equations for the electron component. 
	The ions are assumed to be at rest. Assume for the sake of simplicity that the laser pulse is given, 
	i. e. it propagates in a plasma with a constant speed $v_g$  without changing its shape. 
	The plasma oscillations formed in a wake behind the laser pulse have the form 
	of a wave in which all the functions depend on variables $x$  and $t$  in the combination 
	$\zeta=x-v_{ph} t$, i. e. the wake wave propagates with the constant phase velocity $v_{ph}$  
	equal to the group velocity of the driver laser pulse,  $v_g$. 
	The longitudinal component of the electron momentum obeys the nonlinear differential equation 
	{(e. g. see Ref. \cite{Panch})}
\begin{equation}
\left(\gamma_e-\beta_{ph}\frac{p_{||}}{m_ec}\right)^{\prime \prime}
=k_p^2
\left(\frac{p_{||}}{\gamma_em_ec\beta_{ph}-p_{||}}\right). 
\label{eq:pWaWa}
 \end{equation}
Here a prime denotes differentiation with respect to the variable  $\zeta$. 
The electron relativistic Lorentz factor  $\gamma_e$ depends on the longitudinal and transverse (with respect to the wave propagation direction)
components of the electron momentum,  $p_{||}$ and $m_eca_0$  as  
$\gamma_e=\sqrt{1+a_0^2+(p_{||}/m_ec)^2}$. 
Here it is taken into account that due to the homogeneity
along the transverse direction  
of the configuration under consideration, 
the transverse component of electron momentum is proportional to the corresponding component 
of the normalized vector potential  $a_0$. Eq. (\ref{eq:pWaWa}) becomes singular 
when the denominator in the term on the right hand side vanishes. The singularity corresponds 
to the case when the electron velocity $p_{||}/m_e \gamma_e$  
becomes equal to the wave phase velocity  $c\beta_{ph}$, 
which means the wave-breaking. In a stationary wave the singularity 
occurs at the maximum of the electron velocity  
$v_m=p_{||,m}/m_e \gamma_{e,m}$. 
We denote the wave-breaking coordinate as $\zeta_m$. 

In order to find the structure of the singularity, we expand the electron momentum 
and Lorentz factor near the singularity at  $\delta \zeta = \zeta - \zeta_m \to 0$, assuming smallness of  
$\delta p_{||}=p_{||,m}-p_{||}$, where  $p_{||,m}=\beta_{ph}\sqrt{(1+a_m^2)/(1-\beta_{ph}^2)}$,  
$\delta p_{||}/p_{||,m} \ll 1$. Here $a_m=a_0(\zeta_m)$  is the vector potential value at the wavebreaking point. 
Keeping the main terms on both sides of Eq. (\ref{eq:pWaWa}), we obtain 
\begin{equation}
\delta p_{||}=-m_ec\beta_{ph}\gamma_{ph}^3\left[\frac{3\sqrt{1+a_m^2}}{2\beta_{ph}}k_p\delta \zeta \right]^{2/3} .
\label{eq:deltaPsing}
 \end{equation}
For the electron velocity we have 
\begin{equation}
v_{el}=
c\beta_{ph}-c \frac{\beta_{ph}}{\gamma_{ph}}\left[\frac{3\sqrt{1+a_m^2}}{2\beta_{ph}}k_p\delta \zeta \right]^{2/3}. 
\label{eq:velsing}
 \end{equation}
This type of singularity in the catastrophe theory is called the "cusp catastrophe" \cite{PSARNO}. 
In local coordinates it can be written as a mapping $\{x\}\to \{y\}$: $y_1=x_1$,  $y_2=x^3_2+x_1x_2$. 

Although in the vicinty of the singularity the electron density tends to infinity as 
\begin{equation}
n=\frac{n_0c\beta_{ph}}{c\beta_{ph}-v_{el}}\approx n_0 \gamma_{ph}
\left[\frac{3\sqrt{1+a_m^2}\beta_{ph}}{2 k_p\delta \zeta} \right]^{2/3},
\label{eq:denssing}
 \end{equation}
the singularity is integrable, i.e. the breaking plasma wave contains a finite number of particles. 

\subsection{Above-Barrier Reflection from Caustics Formed by Breaking Plasma Waves}

The plasma wave at the wave breaking threshold can reflect a counter-propagating 
electromagnetic wave due to above-barrier reflection. 
Following the widely used terminology, we refer to the singularities 
in the electron density formed during the plasma wave breaking 
(singularities of Lagrangian mapping) as the caustics. 

In order to calculate the reflection and transmission coefficients
 we represent electromagnetic wave in Eqs. (\ref{eq:wavezeta}) and  (\ref{eq:qzeta}) in the form (see Refs. \cite{Panch, THERMO-II})
\begin{equation}
a(\zeta)=\frac{1}{\sqrt{q(\zeta)}}\left[b_{+}\exp (iW(\zeta))+ b_{-}\exp(-iW(\zeta))\right],
\label{eq:a+-waves}
 \end{equation}
 where the phase integral $W(\zeta)$ is given by
\begin{equation}
W(\zeta)=\int^{\zeta}_0 q(z) dz.
\label{eq:Wzeta}
 \end{equation}

\begin{figure}
	    \begin{center}
 \begin{tabular}{c}
    \includegraphics[
	width=9cm,height=4.5cm
	]{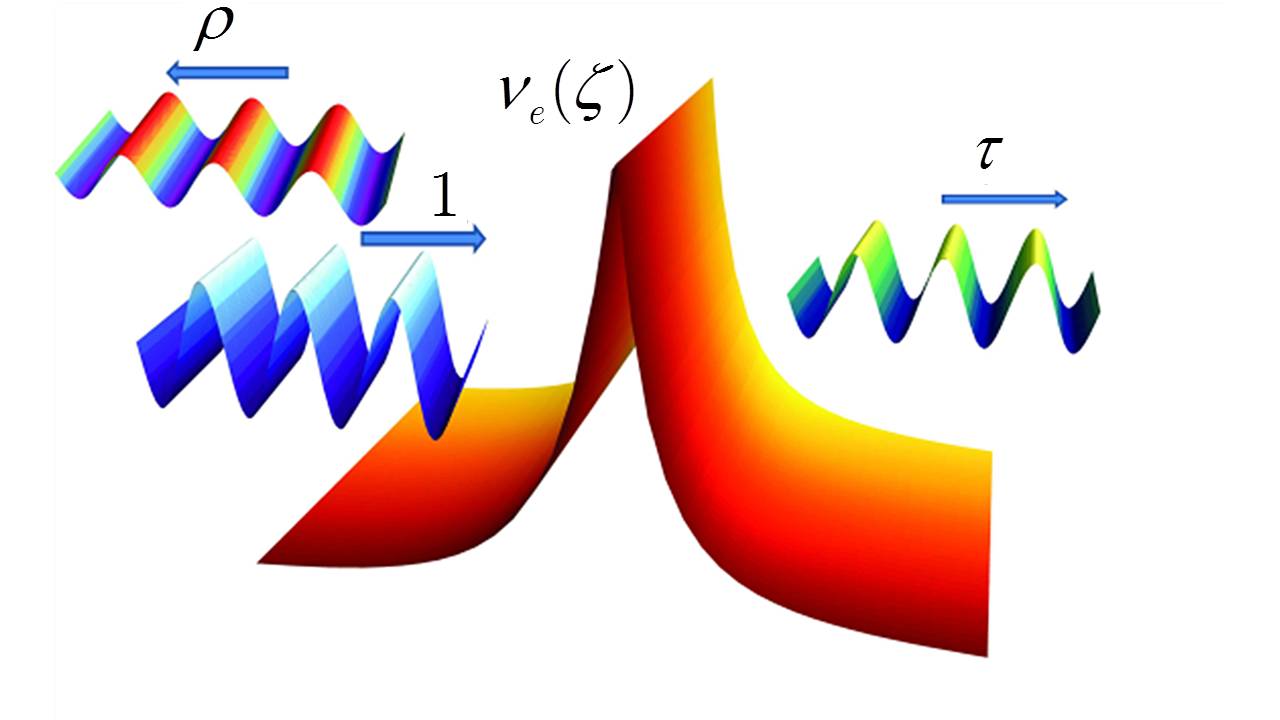}
	\end{tabular}
	\end{center}
	\caption{ Partial reflection of the electromagnetic wave
incident on the moving density cusp \cite{Panch, THERMO-II}. 
}
		\label{RT-CAUSTICS}
	\end{figure}

 The functions $b_{+}$  and  $b_{-}$ are related to the amplitudes of the reflected and transmitted waves.
 In the limit $\zeta \to -\infty$  the function $b_{+}$  equals the amplitude of the incident wave, 
 which is assumed equal to unity, 
 and  $b_{-}(-\infty)=\rho$ corresponds to the amplitude of the reflected wave. 
 For $\zeta \to +\infty$  the function $b_{+}$  equals the amplitude of the transmitted wave $\tau$, 
 and  $b_{-}$ vanishes (see Fig. \ref{RT-CAUSTICS}). 
 Therefore $|b_{+}(-\infty)|^2=1$, $|b_{-}(-\infty)|^2={\rm {\bf R}}$, 
 $|b_{+}(+\infty)|^2={\rm {\bf T}}$ and  $|b_{-}(+\infty)|^2=0$.

 Since in the representation (\ref{eq:a+-waves}), the single unknown
function $a(\zeta)$ has been replaced by the two unknown functions
$b_{+}$ and $b_{-}$, an additional condition is necessary. We 
impose the condition
\begin{equation}
\frac{d a}{d \zeta}=i\sqrt{q(\zeta)}\left[b_{+}\exp (iW(\zeta))- b_{-}\exp(-iW(\zeta))\right],
\label{eq:da-dzeta}
 \end{equation}
 
 Eqs. (\ref{eq:a+-waves}) and (\ref{eq:da-dzeta}) result in the system of equations which can be written as
\begin{equation}
\frac{d}{dW}\left( 
\begin{array}{c}
b_{+} \\ 
b_{-}%
\end{array}%
\right) =\left( 
\begin{array}{cc}
0 & S(W)\exp (-2iW) \\ 
S(W)\exp (2iW) & 0%
\end{array}%
\right) \left( 
\begin{array}{c}
b_{+} \\ 
b_{-}%
\end{array}%
\right) 
\label{eq:dbpm/dW}
 \end{equation} 
 with
\begin{equation}
S(W)=\frac{1}{2}\frac{d}{dW}{\rm ln}q(\zeta(W)).
\label{eq:dadzeta}
 \end{equation}
 
 Integrating both sides of Eq. (\ref{eq:dbpm/dW}) and using the above
formulated boundary conditions for $b_{+}$ and $b_{-}$, one can obtain
the reflection coefficient $\rho$ in the form of an infinite
series \cite{BBB}. In the case of weak reflection, $|\rho| \ll 1$, which requires
$s^2 \gg \nu_e(\zeta)$, the reflected wave can be found
within the framework of perturbation theory.  This yields
\begin{equation}
\rho=\frac{i}{2 s}\int^{+\infty}_{-\infty}\nu_e(\zeta)\exp(-2is\zeta)d\zeta.
\label{eq:rho-s}
 \end{equation}
 
 As an important example of the calculation of the reflection coefficient,
we consider a typical dependence of the electron density
in the vicinity of the wavebreaking threshold in a cold
plasma. This can be approximated by the expression \cite{THERMO-II}
\begin{equation}
\nu_e (\zeta)=\frac{g_{2/3}}{(k_p l)^{2/3}(l^2+\zeta^2)^{1/3}}.
\label{eq:nu-s}
 \end{equation}
Here $g_{2/3}=(2/9)^{1/3}(1+a_m^2)^{1/6}k_p^{4/3}\gamma_m^{4/3}$ is the dimensionless coefficient and the parameter
$k_p l$ shows how close the wave is to the wavebreaking
limit, for which $l=0$. Calculating the integral (\ref{eq:rho-s}) for the integrand (\ref{eq:nu-s}), we find
\begin{equation}
\rho_{2/3}(s,l)=\frac{3 i \sqrt{\pi}g_{2/3}l^{1/6}}{\Gamma(-2/3)k_p^{2/3}s^{7/6}}K_{1/6}(2sl).
\label{eq:rho2/3}
 \end{equation}
Here $\Gamma(x)$ and $K_{\nu}(x)$ are the Euler gamma function and the
modified Bessel function, respectively.

In the limit of relatively large $l$, when $s l  \gg  1$, the
above-barrier reflection is exponentially weak,
\begin{equation}
\rho_{2/3}(s,l)\approx\frac{3 i \pi g_{2/3}}{\Gamma(-2/3)k_p^{2/3}s^{5/3}l^{1/3}}\exp (-2sl).
\label{eq:rho2/3exp}
 \end{equation}
In the opposite case, when $sl \to 0$, we have a 
non-exponentially-small reflection 
\cite{Panch, THERMO-II}
\begin{equation}
\rho_{2/3}(s)\approx\frac{3 i (-2)^{2/3} \pi  g_{2/3}}{k_p^{2/3}s ^{4/3} \Gamma(-1/3)}.
\label{eq:rho2/3l0}
 \end{equation}
This yields the reflection coefficient
\begin{equation}
{\rm {\bf R}}_{2/3}(s)\approx\frac{2^{4/3} 9 g^2_{2/3}}{k_p^{4/3}s^{8/3}\Gamma^2(-1/3)}.
\label{eq:R2/3}
 \end{equation}

\section{Compact Source of High-Brightness X-Rays Based on the Mechanism of a Relativistic Flying Mirror}

\subsection{The relativistic flying mirror in the nonlinear wake waves}

\begin{figure}
	    \begin{center}
 \begin{tabular}{c}
    \includegraphics[
	width=15cm,height=10cm
	]{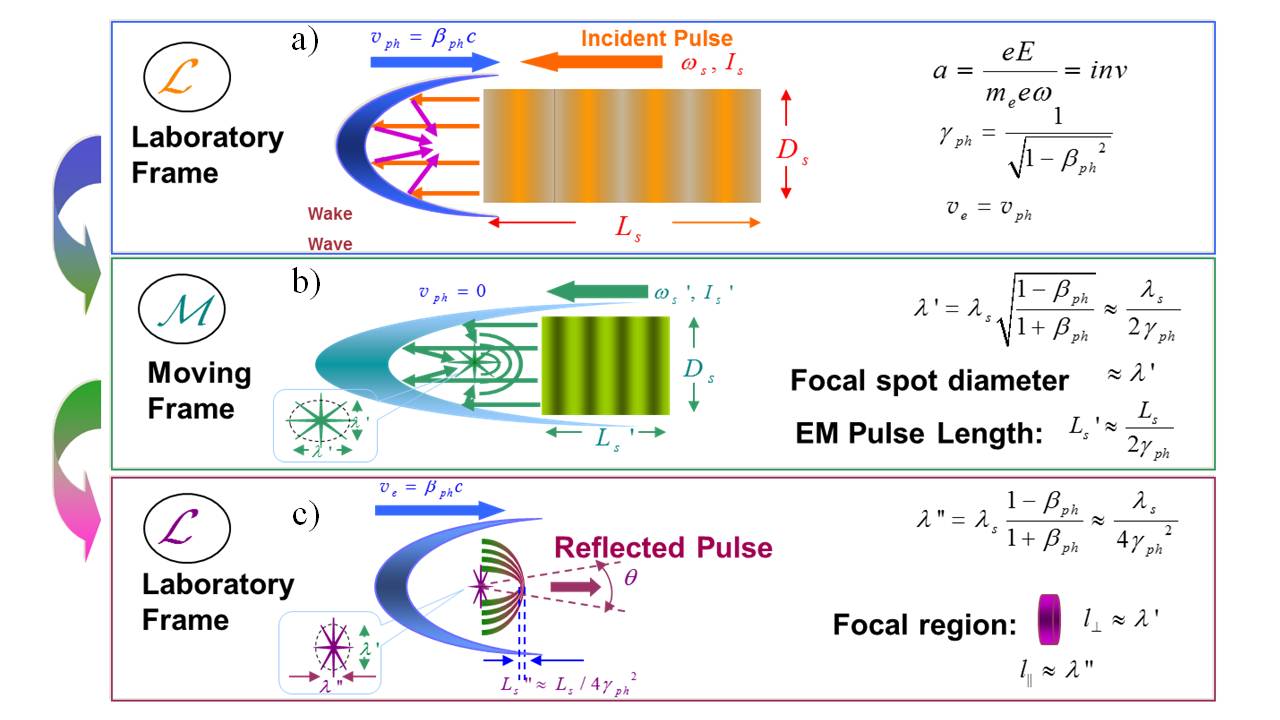}
	\end{tabular}
	\end{center}
		\caption{ The concept of flying relativistic mirror. a) Laboratory frame of reference. 
	Before reflection from the mirror the laser pulse propagates from right to left. 
	b) Frame of reference where the mirror is at rest. 
	The incident laser pulse has a length $L^{\prime}\approx L_s/2\gamma_M$  and wave length  
	$\lambda^{\prime}\approx \lambda_s/2\gamma_M$. 
	After reflection from a parabolic mirror the radiation is focused to the region of size  $\lambda^{\prime}$. 
   c) Laboratory frame of reference. The reflected pulse is compressed by a factor $4\gamma^2_M$; 
	its wavelength becomes equal  $\lambda^{\prime \prime}\approx \lambda/4\gamma_M$. 
	The moving focal region has the form of an ellipsoid 
	with the longitudinal  $\lambda^{\prime \prime}$ and transverse  $\lambda^{\prime \prime}$ dimension, 
	respectively. 
	Radiation is collimated within the angle $\theta \approx 1/\gamma_M$. 
		\label{FIG-MIRR}
		}
	\end{figure}

Within the framework of the RFM concept (see Refs. \cite{BET-03, MIRR} 
and Fig. \ref{FIG-MIRR}) 
the relativistic mirrors are related to the thin layers of relativistic electrons generated 
in nonlinear plasma waves which are excited in the plasma by a short strong laser pulse. 
Surfaces of constant density in the wake of nonlinear waves have the form of paraboloids of revolution \cite{SVBASS, TWB, PMtV, MATLIS, BUCK}. 
The counter-propagating laser pulse is partially reflected from 
these structures. 
This results in a frequency upshifting of the reflected pulse, its shortening in the longitudinal 
direction and focusing 
in the transverse direction, providing the intensification 
of reflected radiation, despite the fact that the reflection coefficient is relatively small 
(see results of PIC simulations \cite{BET-03}).

	The relativistic flying mirror with large enough reflection coefficient of the counter-propagating radiation, 
	is formed in the process of the wake wave breaking. The phase velocity of the wake wave $v_{ph}$  equals 
	the group velocity of the driver laser pulse  $v_g=\partial \omega/\partial k$, 
	which, in accordance with the dispersion 
	equation for the transverse (electromagnetic) waves,  $\omega=\sqrt{k^2c^2+\omega^2_{pe}}$, 
	is given by formula  $v_g=c\sqrt{1-(\omega_{pe}/\omega)^2}$. 
	In a low density plasma, where the laser frequency is much higher than the Langmuir 
	frequency  $\omega \gg \omega_{pe}$, 
	the group velocity of electromagnetic waves is close to the speed of light in vacuum. 
	At the threshold of wave breaking the velocity of the electrons 
	in the wake wave becomes comparable in magnitude with the wake wave phase velocity  
	$v_{ph}$, which is equivalent to 
	the condition for the electron energy ${\cal E}_e=m_ec^2\gamma_e$,
where  $\gamma_e$ is the Lorentz factor calculated for  $v_e=v_{ph}$. 
	The wave breaking occurs if \cite{AP}
\begin{equation}
\gamma_e\geq\gamma_{ph}=(1-v_{ph}^2/c^2)^{-1/2}=\omega/\omega_{pe}.
\label{eq:breakcond}
 \end{equation}

As a result of the plasma wave breaking a singularity in the distribution of electron density occurs, 
which breaks the geometrical optics approximation and increases the reflection coefficient. 
If there is no wave-breaking, the reflectivity is exponentially weak (see Eq. (\ref{eq:rho2/3exp})).

In the model, when the singularity is approximated by a cusp, the reflection coefficient is given 
by the formula (\ref{eq:R2/3}). 
In terms of the number of photons, it depends on the density of the plasma as  
$R_{2/3}\approx 1/\gamma_{ph}^4=(\omega_{pe}/\omega)^4=(n/n_{cr})^2$ in the limit  $\gamma_{ph} \gg 1$. 
Taking into account the pulse compression in the transverse direction as a result of focusing 
by the parabolic mirror of the sourse pulse with the intensity $I_s$, diameter  $D_s$, 
and wavelength $\lambda_s$, 
we obtain for reflected wave intensity
\begin{equation}
I_r\approx I_s\left(\frac{D_s}{\lambda_s}\right)^2\gamma_{ph}^2.
\label{eq:IntMirr}
 \end{equation}
 The reflected wave energy and power equal to ${\cal E}_r\approx {\cal E}_s/\gamma_{ph}^2$  
 and  ${\cal P}_r\approx {\cal P}_s$, respectively.
 
 As an example, let's consider the parameters required to achieve the limit when the electron-positron pairs 
 can be created from vacuum \cite{SSB-PRL2010}, 
 $I_r\approx 2\times 10^{27}{\rm W/cm}^2$,
 for the electromagnetic wave reflected by a flying mirror. We assume that the source pulse has the power 
equal to 4 PW. For $\lambda_s=1\mu$m we obtain from Eq. (\ref{eq:IntMirr}) that $\gamma_{ph}=60$. The 
wake wave breaking condition (\ref{eq:breakcond}) yields the driver laser pulse amplitude $a_d=11$, i.e. for the
driver pulse intensity we have $I_d\approx 1.5\times 10^{20}{\rm W/cm}^2$. Since the gamma factor $\gamma_{ph}$ 
associated with the wake wave phase velocity is equal to $\sqrt{n_{cr} a_d/n_e}$ we find from (\ref{eq:breakcond})
 that the plasma density is $\approx 3.3\times 10^{17}{\rm cm}^{-3}$. In order to excite the wake wave in the plasma 
 with such density the laser pulse length should be equal to $l_{las}=\lambda_d \sqrt{n_{cr} a_d/4 n_e}$, i.e. 
 $l_{las}\approx 30\mu$m or the laser pulse duration of $\tau_{las}\approx 100$fs. 
 Assuming the laser pulse driver diameter to be of the order of $l_{las}$ 
 we find the driver pulse power and energy 
 requal to ${\cal P}_{las}=15$PW and ${\cal E}_{las}=1.5$kJ, respectively.

The use of spherical Langmuir waves as a relativistic mirror can enable 
a much higher intensification 
 of the reflected radiation \cite{SSB-SphW}. Owing to the spherical geometry, the reflected wave with 
 a wavelength $\approx \lambda_s/4\gamma_{M}^2$  is focused in the volume  $\approx \lambda_s^3/48\gamma_{M}^6$, 
 that gives for the radiation intensity $I_r\approx I_s \gamma_{M}^4$  for the reflectron coefficient  
 $R_{2/3}\approx 1/\gamma_{M}^4$. {This has important implementations for studying the feasibility of experiments 
 on the photon-photon scattering in vacuum \cite{JKKOGA}.}
 
\section{Experimental Demonstration of a Relativistic Flying Mirror}

The proof-of-principle experiments of the RFM concept have been done in Refs. \cite{K1,P1}. 
There the generation of soft X-rays with a narrow energy spectrum has been reported. 
In this experiment, 
two short laser pulses collided at an angle of 135$^o$ 
in a supersonic gas jet. 
The laser used in this experiment generates pulses of radiation having a wavelength 820 nm, 
with the energy of 210 mJ, with duration of 76 fs, i. e. it has the power of 2.75 TW. 
The second pulse (the source pulse) is focused 
 providing a spatial overlay with the first pulse-driver. 
 The beam collision configuration is not head-on in order to avoid the laser damage 
 by the radiation propagating in the opposite direction relative to the pulse driver. 
 The driver pulse has the intensity of $5\times 10^{17}$ W/cm$^2$. 
The source pulse intensity equals  $\approx 10^{17}$ W/cm$^2$.

In this configuration 
the frequency of light reflected from the relativistic mirror should be
increased by a factor  
$\omega_r/\omega_s\approx 3.4\gamma_{ph}^2$. 
The plasma density in the target is approximately equal to  $5\times 10^{19}$cm$^{-3}$. 
For this plasma density the wake wave gamma factor is $\gamma_{ph}= 6.5$ and 
the theoretically predicted frequency upshifting factor is 140.

For the above written parameters of the laser radiation and the target plasma, 
the laser pulse driver excites the wake wave with a large enough amplitude 
exceeding the wave-breaking threshold. However, the excess over the wavebreaking 
threshold is not too great, and the regular wake wave structure is not destroyed. 
Evidence of this regime is the observation of quasi-mono-energetic spectra 
of electrons accelerated by the wake wave with ultrarelativistic energy of 20 MeV. 
Another indication of the nonlinear character of the wake is the detection 
of upshifted and downshifted maxima in the frequency spectrum of 
the  laterally scattered  radiation corresponding to the stimulated 
backward Raman scattering. In addition, an analysis of 
interferogram reveals the channel formation in the plasma density, 
which is also in accordance with the wake excitation by the 
laser pulse undergoing the relativistic self-focusing leading 
to redistribution of the electron density.

In the experiments \cite{K1,P1}, 
the detected frequency multiplication factor is in the range from 55 to 114, 
corresponding to a reflected radiation wavelength from 7 to 15 nm.

Further development of the theory of relativistic flying mirror and 
further experiments a higher power laser have allowed to demonstrate high-efficiency regimes
\cite{K2}, 
which provides opportunities for developing highly efficient sources in the energy 
range corresponding to hard X-rays. 
In Ref. \cite{K2}, a high reflection reaching the theoretical prediction
has been demonstrated using the laser with an energy of 0.5 J and a power of 15 TW.
In contrast to previous experiments, 
special efforts enabled head-on collision of the beams,
which secured a three orders of magnitude higher number of reflected photons in the spectral range corresponding to soft X-rays. 
To detect the reflected radiation 
the used spectrometer had a relatively large aperture for collecting 
a large enough number of photons. New techniques have been implemented to perform pulse collision at 
a predetermined place with high accuracy at the desired time.

The number of photons of hard electromagnetic radiation, registered in the experiment \cite{K2} 
allows us to conclude that 
the agreement with theoretical predictions for the reflection coefficient, 
defined by the formula (\ref{eq:R2/3}), is achieved.

In addition to the question on the relativistic mirror reflectivity 
it is of particular interest to determine whether the mirror 
has a high quality reflective surface, that is smooth enough. 
In the experiment presented in Ref. \cite{K2} the observed 
radiation in the wavelength range from 12 nm to 20 nm within the observation angles 
in the range from $9^o$ to $17^o$ has a sufficiently smooth spectral distribution. 
Since the frequency upshifting factor depends on the angle at which the 
reflected light propagates and on the local angle of reflection from the mirror surface, 
this shows that the radiation is reflected from a smooth surface having a curvature. 
This is also consistent with the predictions of the theory, which is important for further research 
aimed at demonstrating the sharp focusing of X-rays in order to reach the high limits of its intensity.

\section{Extremely High Field Limits}

The purpose of further studies of the physical processes
associated with relativistic mirrors is to design and build a
compact source of hard electromagnetic radiation with a
photon energy and intensity large enough to conduct
experiments in previously inaccessible regimes of interaction
of electromagnetic fields with matter. Here, we
present the critical parameters that characterize the 
main regimes of
laser-matter interaction, depending on the intensity of the
electromagnetic wave (see also reviews \cite{MTB, MIRR, NIMA, MaShu-2006, EPS-Stra, ADiP-2012}).

In the limit of extremely high wave intensity  
the radiation friction effects on the charged particle dynamics become dominant \cite{RAD1, RAD2, RAD3, JKK, AGZSM, Thomas}. 
At these limits the electron dynamics become dissipative with
fast conversion of the electromagnetic  wave energy to hard electromagnetic radiation, which for
typical laser parameters is in the gamma-ray range \cite{Ridgers, NakaKo}.
For laser radiation with $1\,\mu$m wavelength the radiation friction force changes the scenario
of the electromagnetic wave interaction with matter at the intensity of about $I_R\approx 10^{23}$W/cm$^{2}$. 
For the laser intensity close to $I_R$ also novel physics of abundant electron-positron
pair creation comes into play \cite{BELL} (see \cite{Fedotov} 
and \cite{SSB, NIMA}). In this regime, the electron (positron) interaction with the electromagnetic field is
principally determined by a counterplay between the radiation friction and quantum effects. 
The quantum electrodynamics (QED) effects weaken the electromagnetic 
emission by the relativistic electron resulting in 
the lowering of the radiation friction \cite{JSCHW}. In an extremely high intensity electromagnetic field, vacuum 
shows various nonlinear quantum electrodynamics 
processes such as 
vacuum polarization, electron-positron pair plasma creation, 
and other properties depending on the electromagnetic field 
strength and distribution. 

The probabilities of the processes involving extremely high intensity electromagnetic field interaction with 
electrons, positrons and photons are determined by several dimensionless parameters.

We consider the nonlinear electrodynamics
of plasma in the limit of relativistic particle energies. As was mentioned above, the
behavior of an electron in the field of an electromagnetic wave
is determined by the dimensionless parameter $a=eE/m_e \omega c$,
which is the normalized wave amplitude. This value is
associated with a relativistic invariant, 
\begin{equation}
a=e\frac{\sqrt{A^{\mu}A_{\mu}}}{m_e c^2}.
\label{eq:a-dimless}
 \end{equation}
If $a > 1$, the electron energy becomes relativistic, which corresponds to
the wave intensity above $\approx 1.37\times 10^{18} W/cm^2$.

With further increase in the electromagnetic wave
intensity, its interaction with a sufficiently dense plasma is
characterized by radiation loss effects \cite{RAD1, RAD2, RAD3}. The energy
loss rate by an electron rotating in the field of a circularly
polarized wave is given by
\begin{equation}
\dot{\cal E}^{(-)}=\epsilon_{rad} m_e c^2 \omega a^2 (1+a^2).
\label{eq:rad1}
 \end{equation}
Here the dimensionless parameter 
\begin{equation}
\epsilon_{rad} = \kappa_p \frac{2\pi r_e}{\lambda}
\label{eq:e-rad}
 \end{equation}
characterizes the the radiation losses with $\lambda=2\pi c/\omega$ and the classical 
electron radius equal to $r_e=e^2/m_e c^2\approx 2.8\times 10^{-13} cm$. 
The coefficient $ \kappa_p$ for circularly polarized wave equals 3/8 and for linear 
polarization it is 1/8.

Because the electromagnetic wave cannot transfer energy
to the particle with a rate greater than
\begin{equation}
\dot{\cal E}^{(+)}=e E c= m_e c^2 \omega a
\label{eq:rad2}
 \end{equation}
the condition of the balance between incoming and
radiated energy per unit time, $\dot{\cal E}^{(+)}=\dot{\cal E}^{(-)}$
yields the
threshold value of the amplitude of the electromagnetic
wave above which the influence of radiation losses cannot
be ignored. In dimensionless form, this threshold amplitude of
the wave is
\begin{equation}
a_{rad}=\epsilon_{rad}^{-1/3}.
\label{eq:rad3}
 \end{equation}
For a circularly polarized wave this corresponds to the radiation
intensity 
\begin{equation}
I_R=\left(\frac{3}{2}\right)^{2/3}\frac{m_e^{8/3} c^5 \omega^{4/3}}{4 \pi e^{10/3}}
=2.65 \times 10^{23} \left(\frac{1 \mu {\rm m}}{\lambda}\right)^{4/3}\frac{\rm W}{{\rm cm}^2}.
\label{eq:Irad}
\end{equation}

QED effects become important, when
the energy of a photon generated by Thomson (Compton) scattering is of
the order of the electron energy, i.e. $\hbar \omega_m \approx m_e c^2 \gamma_e$. 
If $\gamma_e=a_0$ this yields the quantum electrodynamics limit on the electromagnetic field amplitude, $a_0^2/a_S>1$. Here 
the dimensionless parameter
\begin{equation}
a_S=\frac{e E_S \lambda}{2 \pi m_e c^2}=\frac{m_e c^2}{\hbar \omega}=\frac{\lambda}{\lambda_C}=4.2\times 10^{5}\left(\frac{\lambda}{1 \mu {\rm m}}\right)
\label{eq:aS}
\end{equation}
is the normalized critical electric field of quantum electrodynamics  \cite{BLP}, 
\begin{equation}
E_S=m_e^2 c^3/e \hbar,
\label{eq:EScrit}
\end{equation}
with $\lambda_C=2 \pi \hbar /m_e c=2.42\times 10^{-10}$cm 
being the Compton wavelength. 
The electromagnetic radiation intensity (irradiance) corresponding to 
the wave amplitude of $a_S$ is 
\begin{equation}
I_S=\frac{m_e^4 c^7}{4 \pi e^2 \hbar^2}=2.36 \times 10^{29} 
\frac{\rm  W}{{\rm cm}^2}.
\label{eq:IS}
\end{equation}
The above obtained  quantum electrodynamics limit, $a_0^2/a_S>1$, corresponds to the condition $\chi_e >1$, 
where the relativistic and gauge invariant parameter $\chi_e$,
\begin{equation}
\chi_e=\frac{\sqrt{\left(F^{\mu \nu} p_{\nu}\right)^2}}{E_S m_e c }\approx 2\frac{a_0}{a_S}\gamma_e ,
\label{eq:chie}
\end{equation}
characterizes the probability of the gamma-photon emission 
by the electron with 4-momentum $p_{\nu}$ in the field of the electromagnetic wave. 
The 4-tensor of the electromagnetic field is defined as $F_{\mu \nu}=\partial_{\mu} A_{\nu}-\partial_{\nu} A_{\mu}$. 
The quantum electrodynamics limit is reached for 
the electromagnetic wave intensity of the order of 
\begin{equation}
I_Q=\frac{m_e^3 c^5 \omega}{8 \pi e^2 \hbar}=5.75 \times 10^{23} 
\left(
\frac{1 \mu {\rm m}}
{\lambda}
\right)
\frac{\rm  W}{{\rm cm}^2}.
\label{eq:IQ}
\end{equation}

The Lorentz invariant dimensionless parameter 
\begin{equation}
\chi_{\gamma}=\frac{\hbar \sqrt{\left(F^{\mu \nu} k_{\nu}\right)^2}}{E_S m_e c }
\approx 2\frac{a_0}{a_S^2}\frac{\omega_{\gamma}}{\omega}
\label{eq:chig}
\end{equation}
determines the probability of the electron-positron 
pair creation by the photon with the energy $\hbar \omega_{\gamma}$ in the electromagnetic field via the
Breit-Wheeler process \cite{BW, NR}.

Properties of the nonlinear quantum vacuum  are determined by the Poincare invariants, 
\begin{equation}
\mathfrak{F}=({\bf B}^2-{\bf E}^2)/2 \quad {\rm and} \quad \mathfrak{G}=({\bf E \cdot B}). 
\end{equation}
The probability 
of the electron-positron pair creation depends on 
\begin{equation}
\mathfrak{E}=\sqrt{\sqrt{\mathfrak{F}^2+\mathfrak{G}^2}-\mathfrak{F}} 
\quad {\rm and} \quad 
\mathfrak{B}=\sqrt{\sqrt{\mathfrak{F}^2+\mathfrak{G}^2}+\mathfrak{F}},
\end{equation}
which are the electric and magnetic fields in the frame of reference where they are 
parallel.

In the limit $\mathfrak{E}/E_S \to 1$ 
the electromagnetic waves can create electron-positron pairs from vacuum \cite{H-E, SCHWINGER, BLP, VR, POPOV}, 
moreover, the electromagnetic waves can interact via photon-photon collisions \cite{JKK} 
and vacuum polarization \cite{NNR}. When the parameter $\chi_e$ becomes 
large, the multiphoton Compton scattering results in the high 
energy photon generation. For $\chi_{\gamma} > 1$ the multiphoton
Breit-Wheeler process results in electron-positron pair generation 
via the gamma-photon interaction with the strong electromagnetic field. 
The nonlinear Thomson scattering  regime 
is realized for $a\gg 1$ with the scattering cross-section depending on the electromagnetic field amplitude. 
If  $a_0>a_{rad}=\varepsilon^{-1/3}_{rad}$ radiation friction effects play a key role.

The comparison of expressions given by Eqs. (\ref{eq:Irad}) and (\ref{eq:IQ}) shows 
that for the EM wave length equal to $\lambda\approx 0.8 \mu {\rm m}$ the intensities 
$I_R$ and $I_Q$ are of the same order of magnitude as has been noted in Refs. \cite{FUNDASTRO, BRADY, TZEPLA, MJIRKA}. 
More precisely, the curves $I_R(\omega)$ and $I_Q(\omega)$ intersect each other at the frequency equal to 
\begin{equation}
\omega_1=\frac{e^4 m_e}{18 \hbar^3}
\label{eq:om1}
\end{equation}
corresponding to the wavelength of $\lambda_1=0.821 \mu$m and the photon energy of the order of 1.5 eV.
This implies that QED effects in laser matter interactions soon start to be
revealed at near-future super-intense laser facilities \cite{ELIILE}.

 \section*{Acknowledgment}
 
  This review article is based on the materials of the plenary lecture presented at the ICPIG-2015 conference (Iasi, Romania, 2015) 
  by S.V.B., who thanks Profs. A. Popa and V. Zamfir for organizing the event.
The lecture and review  present recent results obtained at JAEA.

\end{document}